\documentclass[aps,prr,twocolumn,amsmath,amssymb,groupedaddress,longbibliography
]{revtex4-2}

\usepackage{graphicx}
\usepackage{dcolumn}
\usepackage{bm}
\usepackage{multirow}
\usepackage{braket}
\usepackage{color}
\usepackage{ulem}
\usepackage{here}
\usepackage{times}
\usepackage{booktabs}
\usepackage{comment}

\allowdisplaybreaks[4]

\begin{document}

\title{Predominant Electronic Order Parameter for Structural Chirality \\ -- Role of Spinless Electronic Toroidal Multipoles}

\author{Rikuto Oiwa$^{1}$}
 \email{roiwa@phys.sci.hokudai.ac.jp}
\author{Hiroaki Kusunose$^{2}$}
\affiliation{
$^1$Graduate School of Science, Hokkaido University, Sapporo 060-0810, Japan, \\
$^2$Department of Physics, Meiji University, Kawasaki 214-8571, Japan
}

\begin{abstract}
We discuss predominant order parameters for structural chirality, and demonstrate that time-reversal-even axial-quadrupole plays a key role in stabilizing a chiral structure.
Using the symmetry-adapted closest Wannier model of the trigonal Te and Se, we quantify the evolution of the spin-independent (spinless) and spin-dependent (spinful) electric toroidal (ET) (axial) multipole moments across the transition from an achiral to a chiral structure.
Our results clearly identify that a spin-independent off-diagonal real hopping between $p$ orbitals, which corresponds to the bond-cluster spinless ET quadrupole of $(3z^{2}-r^{2})$ type $G_{u}$, is the predominant order parameter in stabilizing helical structures.
We further elucidate that the above itinerant spinless ET quadrupole induces a monopole-like orbital angular momentum texture in the momentum space, which can be observed via circular-dichroism in soft x-ray photoemission spectroscopy measurement.
Our findings highlight a critical role of the orbital angular momentum in chiral materials rather than less dominant spin angular momentum arising from the relativistic spin-orbit coupling.\end{abstract}

\maketitle

\section{Introduction}
\label{sec:intro}

Chirality is ubiquitous in nature, which breaks all of mirror symmetries as well as spatial inversion symmetry.
This symmetry-broken state is caused by an emergence of a time-reversal $\mathcal{T}$-even and spatial-inversion $\mathcal{P}$-odd pseudoscalar that is expressed by an inner product of vectors with the same $\mathcal{T}$ and different $\mathcal{P}$ parities for example~\cite{barron_2004, barron_2013}.
The chiral spin-orbit coupling (SOC), $\bm{k}\cdot\bm{\sigma}$, is such an example, where crystal momentum $\bm{k}$ and spin $\bm{\sigma}$ of an electron are $\mathcal{T}$-odd polar and axial vectors, giving rise to hedge-hog type spin texture in chiral materials~\cite{Hirayama2015, Sakano2020}.

Such spin-dependent electronic state caused by $\mathcal{T}$-even pseudoscalar is considered as an origin of chirality-induced phenomena such as current-induced optical activity~\cite{1979JETPL}, kinetic magnetoelectric (Edelstein) effect~\cite{Yoda2015, Yoda2018, Shalygin2012, Furukawa2017, Furukawa2021}, rotation-field induced electric polarization~\cite{RO_PRL_2022} (called rotoelectricity in Ref.~\cite{Gopalan_NatMater_2011}), and so on.
For a quantitative understanding of such phenomena, it is quite important to examine an electronic order parameter of chirality which is also responsible for stabilizing a chiral structure.
Once an electronic order parameter is identified quantitatively, it could open a way to control crystal chirality via the most dominant coupling to external fields with $\mathcal{T}$-even pseudoscalar property, such as the optical chirality, Zilch $\bm{E}\cdot(\bm{\nabla}\times\bm{E})+\bm{B}\cdot(\bm{\nabla}\times\bm{B})$~\cite{Lipkin_Zilch_1964, Kibble_Zilch_1965, Proskurin_2017, Hoshino_PRL_2023, SH_Floquet_2024, HK_APL_2024}, and the inner product $\bm{J}\cdot\bm{B}$ of an electric current $\bm{J}$ and a magnetic field $\bm{B}$~\cite{HK_APL_2024}, which is a source of electric magneto-chiral anisotropy~\cite{PORTIGAL_jphys_1971,Rikken_nature_1997,kanazawa_nc_2017,rikken_prb_2019,aoki_prl_2019,ishizuka_nc_2020,Atzori_chirality_2021,Suarez_PRB_2025}, and so on.

Recent extensive investigations on chirality-induced spin selectivity (CISS)~\cite{Naaman_CISS_JPCL_2012, Waldeck_CISS_APL_2021, Evers_CISS_AdvMater_2022, Bloom_CISS_Review_2024} stimulate significance of electron spin in chiral materials.
CISS is characterized by considerably high spin polarization or accumulation through a chiral material in which the sign of the polarized spin depends on the handedness of chiral material. 
These phenomena have been widely observed in molecules~\cite{Ray_CISS_1999, Gohler_CISS_2011, Dor_CISS_2017, Michaeli_CISS_2019, Naaman_CISS_ACS_2020, Suda_CISS_natcommun_2019, Miwa2020, Kato_CISS_PRB_2022, Kondou2022, Miwa_CISS_arxiv_2024}, as well as organic and inorganic crystals~\cite{Inui_CISS_PRL_2020, Nabei_CISS_APL_2020, Shishido_CISS_ARL_2021, Shiota_CISS_PRL_2021, Nakajima_CISS_Nature_2023,Ohe2024}.
Accordingly, the spinful electronic order parameters have been proposed to quantify material chirality.
For instance, recent theoretical studies have attempted to quantify electronic chirality using a spinful quantum mechanical operator with the $\mathcal{T}$-even pseudoscalar symmetry based on DFT calculations~\cite{RO_PRL_2022, Kishine_IJC_2022, AI_JCP_2024, Hoshino_PRL_2023, Miki_arxiv_2024}.

Nevertheless, in considering that the relativistic SOC is the microscopic origin of these spinful quantities, an energy gain from spinful order parameters must be one or two orders of magnitude smaller than that from spinless quantities, such as spin-independent crystalline electric fields and electron hoppings.
With regard to the current induced magnetization in elemental Te, it is suggested that the orbital contribution is more significant than the spin magnetization~\cite{Tsirkin_PRB_2018,Arunesh_npjcm_2022}.
Furthermore, the observation of monopole-like orbital angular momentum (OAM) texture in the momentum space has been reported in the cubic B20 family of chiral crystals in Sohncke space group, such as PdGa, PtGa, CoSi, using circular-dichroism (CD) in photoemission spectroscopy measurements~\cite{Yang_OAM_2023, Brinkman_OAM_2024, Yen_OAM_2024}.
In light of these circumstances, the significance of spinless chiral order parameter with orbital degrees of freedom is realized, and it is necessary to investigate comprehensively a predominant and more suitable order parameter among the class of the electric toroidal (ET) multipoles, which break all of mirror and spatial inversion symmetries.

In this paper, we analyze the evolution of the ET multipoles across the achiral-to-chiral structure change by using the symmetry-adapted closest Wannier model of the trigonal Te and Se.
The results show that the ET monopoles and quadrupoles are finite only when the system is chiral and their signs correspond to the handedness of the crystal.
In particular, we show that an itinerant spinless ET quadrupole $G_u$, corresponding to the spin-independent off-diagonal real hopping between $p$ orbitals, is the predominant order parameter in stabilizing the helical structure.
We further show that this itinerant spinless ET quadrupole is a source of the monopole-like OAM texture in the momentum space, which can be observed using CD in soft x-ray photoemission spectroscopy measurements.

The subsequent sections of this paper are organized as follows.
In Sec.~\ref{sec:ET}, we begin with classifying electronic order parameter candidates for each chiral point group based on the symmetry-adapted multipole theory~\cite{SH_HK_2018, SH_MY_YY_HK_mul_2018, HW_YY_Mul_2018, HK_RO_SH_comp_mul_2020, SH_HK_JPSJ_2024}.
In Sec.~\ref{sec:dft}, we present a comprehensive analysis of changes in the crystal and band structures of Te and Se and discuss the stability mechanism of the chiral structure based on the DFT calculations.
Section \ref{sec:scw} provides a detailed discussion of the symmetry-adapted closest Wannier model used in this study.
The results are discussed in Sec.~\ref{sec:results_discussions}, where we show the evolution of each order parameter across the achiral-to-chiral structure change.
We then elucidate that the spinless ET quadrupole $G_{u}$ is the predominant order parameter in stabilizing the helical crystal structures.
In addition, we show that the $G_{u}$ is the leading origin of the monopole-like OAM texture in the momentum space.
Lastly, Sec.~\ref{sec:conclusion} summarizes our investigations and an outlook for future study.

\section{Electric toroidal multipoles as order parameters for structural chirality}
\label{sec:ET}

\begin{table}[t]
\begin{center}
\caption{ \label{tab_SAMB}
Four types of dipole moments $\bm{X}$ $(X = Q,M,T,G)$ classified by the time-reversal $\mathcal{T}$, spatial-inversion $\mathcal{P}$, and mirror $\mathcal{M}_{\parallel}$ ($\mathcal{M}_{\perp}$) operations parallel (perpendicular) to dipole vector. 
The ET monopole $G_{0}$ and quadrupole $G_{u}$ moments are also shown.
The conjugate physical fields are given in the last column.
}
\begin{ruledtabular}
\begin{tabular}{cccccll}
               & $\mathcal{T}$ & $\mathcal{P}$ & $\mathcal{M}_{\perp}$ & $\mathcal{M}_{\parallel}$ & Expression &  Conjugate field  \\
\hline
$\bm{Q}$ & $+$ & $-$ & $-$ & $+$ & $\bm{T}\times\bm{M}, \bm{Q}\times\bm{G}$ & Electric field $\bm{E}$ \\ 
 		& 	   & 	     &       &       & Position ($\bm{r}$) & \\
 		& 	   & 	     &       &       & Polarization ($\bm{P}$) & \\
$\bm{M}$ & $-$ & $+$ & $+$ & $-$ & $\bm{Q}\times\bm{T}, \bm{G}\times\bm{M}$ & Magnetic field $\bm{B}$ \\ 
		& 	   & 	     &       &       & Orbital ($\bm{l}=\bm{r}\times\bm{p}$) &  \\
		& 	   & 	     &       &       & Spin ($\bm{s}$) & \\
$\bm{T}$ & $-$ & $-$ & $-$ & $+$ & $\bm{Q}\times\bm{M}, \bm{G}\times\bm{T}$ & Electric current $\bm{J}$ \\
$\bm{G}$ & $+$ & $+$ & $+$ & $-$ & $\bm{Q}^{(1)}\times\bm{Q}^{(2)}$  & Rotational distortion \\
		& 	   & 	     &       &       & $\bm{M}^{(1)}\times\bm{M}^{(2)}$  & $\bm{\omega} = \bm{\nabla}\times\bm{u}$ \\
\hline
$G_{0}$  & $+$ & $-$ & $-$ & $-$ & $\bm{Q}\cdot\bm{G}, \bm{T}\cdot\bm{M}$ & Zilch $\bm{E} \cdot (\bm{\nabla} \times \bm{E})$ \\
$G_{u}$  & $+$ & $-$ & $-$ & $-$ & $3Q_{z}G_{z}-\bm{Q}\cdot\bm{G}$  &  $\qquad +\bm{B} \cdot (\bm{\nabla} \times \bm{B})$ \\
		& 	   & 	     &       &       & $3T_{z}M_{z}-\bm{T}\cdot\bm{M}$ & $\bm{J}\cdot {\bm{B}}$ \\
	       & 	   & 	     &       &       & $Q_{x} Q_{yz} - Q_{y} Q_{zx}$ & 
\end{tabular}
\end{ruledtabular}
\end{center}
\end{table}

\begin{figure}[t]
  \begin{center}
  \includegraphics[width=0.95\hsize]{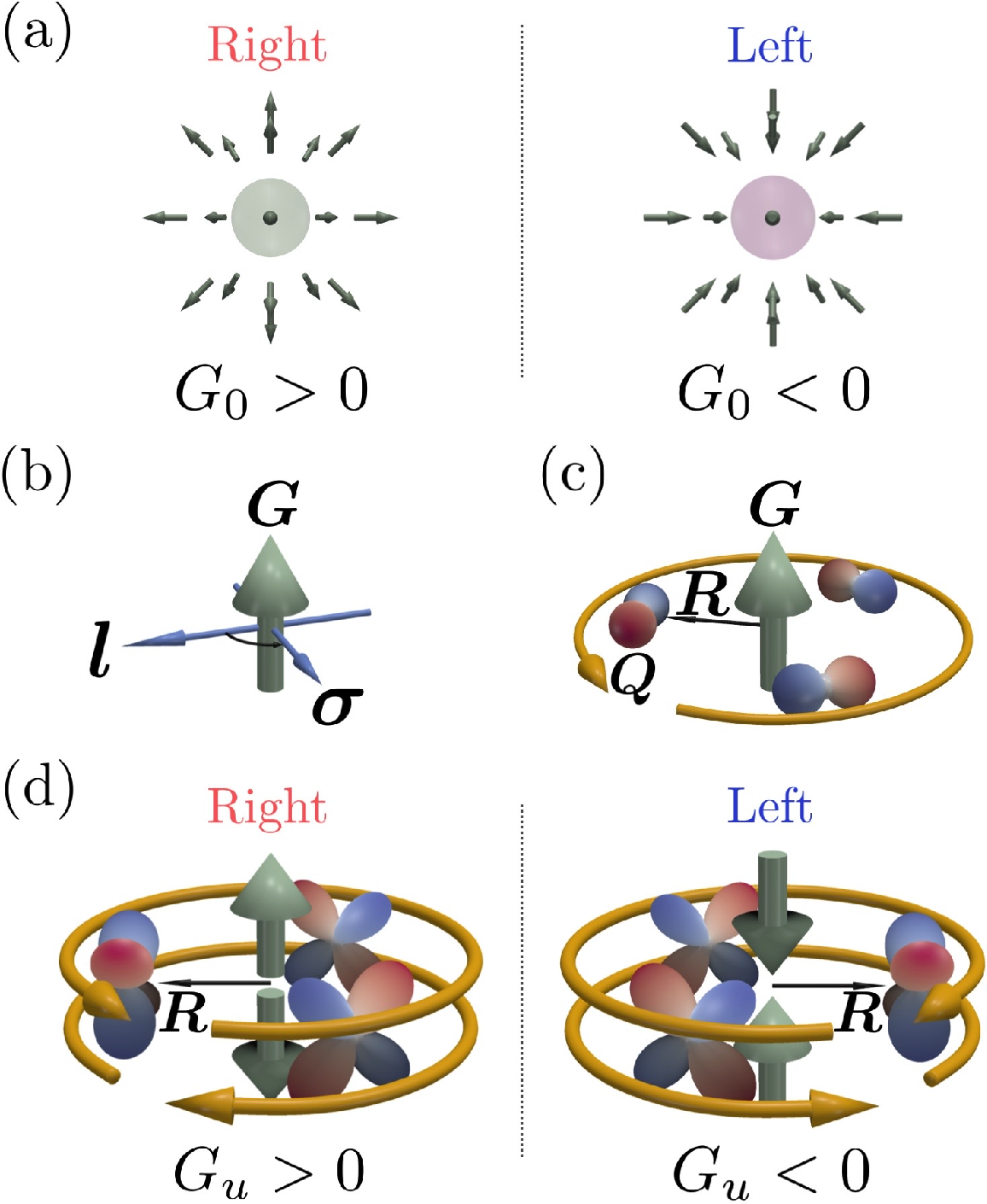}
  \caption{
  \label{fig_ET}
(a) ET monopole $G_{0}$ in terms of the flux structure of the ET dipoles $\bm{G}$, whose divergence corresponds to the handedness.
(b) Atomic ET dipole and (c) site/bond-cluster ET dipole, which is a vortex-like alignment of the E dipoles $\bm{Q}$ at the atoms or bond centers, where the orange vector represents the direction of $\bm{Q}$.
(d) Site/bond-cluster ET quadrupole $G_{u}$, which is a vortex-like alignment of the E quadrupoles, ($Q_{yz}, Q_{zx}$). 
  }
  \end{center}
\end{figure}

An entity of chirality can be discussed unambiguously by using the symmetry-adapted multipole basis (SAMB)~\cite{SH_HK_2018, SH_MY_YY_HK_mul_2018, HW_YY_Mul_2018, HK_RO_SH_comp_mul_2020, SH_HK_JPSJ_2024}.
According to the $\mathcal{P}$ and $\mathcal{T}$ parities, SAMBs are classified into four types: electric (E), magnetic (M), magnetic toroidal (MT), and electric toroidal (ET) multipoles~\cite{SH_HK_2018, SH_MY_YY_HK_mul_2018, HK_RO_SH_comp_mul_2020, SH_HK_JPSJ_2024}.
The E, M, MT, and ET multipoles are further characterized by the rank of multipole $l$ and the irreducible representation (IR) and its component $(\Gamma, \gamma)$ under the point group symmetry of a system, $X_{l\Gamma\gamma}$ ($X = Q, M, T, G$).

Notably, a set of SAMBs $\left\{X_{l\Gamma\gamma}\right\}$ constitutes the symmetry-adapted complete basis set in the Hilbert space of molecules or crystals~\cite{HK_PRB_2023, RO_arxiv_2025}.
Table~\ref{tab_SAMB} shows the symmetry properties of E, M, MT, and ET dipoles ($\bm{Q}, \bm{M}, \bm{T}, \bm{G}$), and the ET monopole $G_{0}$ and quadrupole $G_{u}$ in terms of the parities of $\mathcal{T}$ and $\mathcal{P}$ as well as mirror operations perpendicular (parallel) to dipole vector, $\mathcal{M}_{\perp} (\mathcal{M}_{\parallel})$.

The ET monopole $G_{0}$ with ($\mathcal{T}$, $\mathcal{P}$, $\mathcal{M}_{\perp}$, $\mathcal{M}_{\parallel}) = (+$, $-$, $-$, $-)$ has the same symmetry properties as the $\mathcal{T}$-even pseudoscalar.
Since it only becomes the totally symmetric IR in the 11 chiral point groups: cubic (O, T), triclinic (C$_{1}$), monoclinic (C$_{2}$), orthorhombic (D$_{2}$), tetragonal (C$_{4}$, D$_{4}$), trigonal (C$_{3}$, D$_{3}$), hexagonal (C$_{6}$, D$_{6}$) (see Table XVI in Ref.~\cite{SH_MY_YY_HK_mul_2018}), $G_{0}$ is an order parameter candidate of chirality.
As mentioned above, $G_{0}$ can be expressed by the inner product of polar and axial vectors with the same $\mathcal{T}$ parity, $G_{0} = \bm{Q}\cdot\bm{G}$, or $\bm{T}\cdot\bm{M}$.
Since $\bm{Q}$ is a category of an E dipole, such as position vector $\bm{r}$, $G_{0}$ can be regarded as the flux structure of $\bm{G}$ as shown in Fig.~\ref{fig_ET}(a), and the divergence of the fluxes corresponds to the handedness.

When we consider electronic degrees of freedom within an isolated atom, $G_{0}$ is constructed only with the spin degree of freedom, i.e., it is given by $G_{0}=\bm{r}\cdot(\bm{l}\times\bm{\sigma})$, where $\bm{G} = \bm{l}\times\bm{\sigma}$ is the spinful ET dipole as shown in Fig.~\ref{fig_ET}(b).
When we reexpress $G_{0}$ as $G_{0}=(\bm{r}\times\bm{l})\cdot\bm{\sigma}$, the quantity $\bm{t} = \bm{r}\times\bm{l}$ is the spinless MT dipole.

On the other hand, when we consider the sublattice degrees of freedom, $G_{0}$ can be constructed without spin degree of freedom.
Since the position vector of $i$-th atom, $\bm{R}_{i}$ belongs to the category of $\bm{Q}$, we can also express $G_{0}$ as $\sum_{i}\bm{R}_{i}\cdot\bm{G}_{i}$.
Similarly, given that the real and imaginary hoppings corresponding to E and MT multipoles from the symmetry point of view~\cite{HK_PRB_2023}, the ET monopoles of $(\bm{Q}\cdot\bm{G})_{ij}$ and $(\bm{T}\cdot\bm{M})_{ij}$ ($\bm{M} = \bm{l}, \bm{\sigma}$) can be accomodated in the $(i j)$ bond~\cite{RO_PRL_2022, AI_JCP_2024}, as discussed in more detail in Sec.~\ref{sec:samb}.
The Fourier transform of the latter corresponds to $\bm{k}\cdot\bm{l}$ and $\bm{k}\cdot\bm{\sigma}$, giving rise to the monopole-like OAM~\cite{Yang_OAM_2023, Brinkman_OAM_2024, Yen_OAM_2024} and spin angular momentum~\cite{Hirayama2015, Sakano2020} textures, respectively.

In practical chiral crystals, there are only discrete rotational symmetry.
In such cases, the ET multipoles other than rank $0$ also become totally symmetric IR.
For example, in non-cubic point groups with mono-axial principal axis, the lowest-rank ET multipole is the ET quadrupole of $(3z^{2}-r^{2})$-type $G_{u}$, which has the same $(\mathcal{T}, \mathcal{P}, \mathcal{M}_{\perp}, \mathcal{M}_{\parallel})$ properties as $G_{0}$.
Similarly, in cubic chiral systems with three-fold axes, the ET hexadecapole $G_{4}$ becomes the lowest-rank totally symmetric IR other than $G_{0}$.

As similar to the classical view of $G_{z}$ is a vortex-like alignment of E dipoles, i.e., $G_{z} = \sum_{i} \left[(R_{x})_{i} (Q_{y})_{i} - (R_{y})_{i} (Q_{x})_{i} \right]$, as shown in Fig.~\ref{fig_ET}(c), we can also express $G_{u}$ as 
\begin{align}
G_{u} = Q_{x} Q_{yz}  - Q_{y} Q_{zx}.
\label{eq_Gu_classical}
\end{align}
From symmetry point of view, $\bm{Q}=(Q_{x},Q_{y},Q_{z})$ can be regarded either as the position vector $\bm{R}_{i}$ or the real hopping, and $Q_{zx}$, $Q_{yz}$ can be regarded either as electronic quadrupole moments or real hopping between ($p_{x}, p_{y}$) and $p_{z}$ orbitals.
The schematic picture of Eq.~(\ref{eq_Gu_classical}) is shown in Fig.~\ref{fig_ET}(d), and the divergence of the vorticity corresponds to the handedness.
Notably, $G_{u}$ in Eq.~(\ref{eq_Gu_classical}) is also discussed as an electronic order parameter candidate for the non-magnetic ordered phase of URhSn (See Eq.~(1) in Ref.~\cite{Ishitobi_arxiv_2025}).

In this way, there are several expressions of spinless and spinful $G_{0}$ and $G_{u}$, and all of them are the candidates of the order parameter for structural chirality.
In Sec.~\ref{sec:results_discussions}, we elucidate the predominant order parameter by using the symmetry-adapted closest Wannier model of Te and Se as specific examples.
In these examples, we can quantitatively identify the most dominant contribution of $G_{0}$ or $G_{u}$ in the model Hamiltonian. 

\section{Stability of chiral structures}
\label{sec:dft}

\begin{figure}[t]
  \begin{center}
  \includegraphics[width=0.99\hsize]{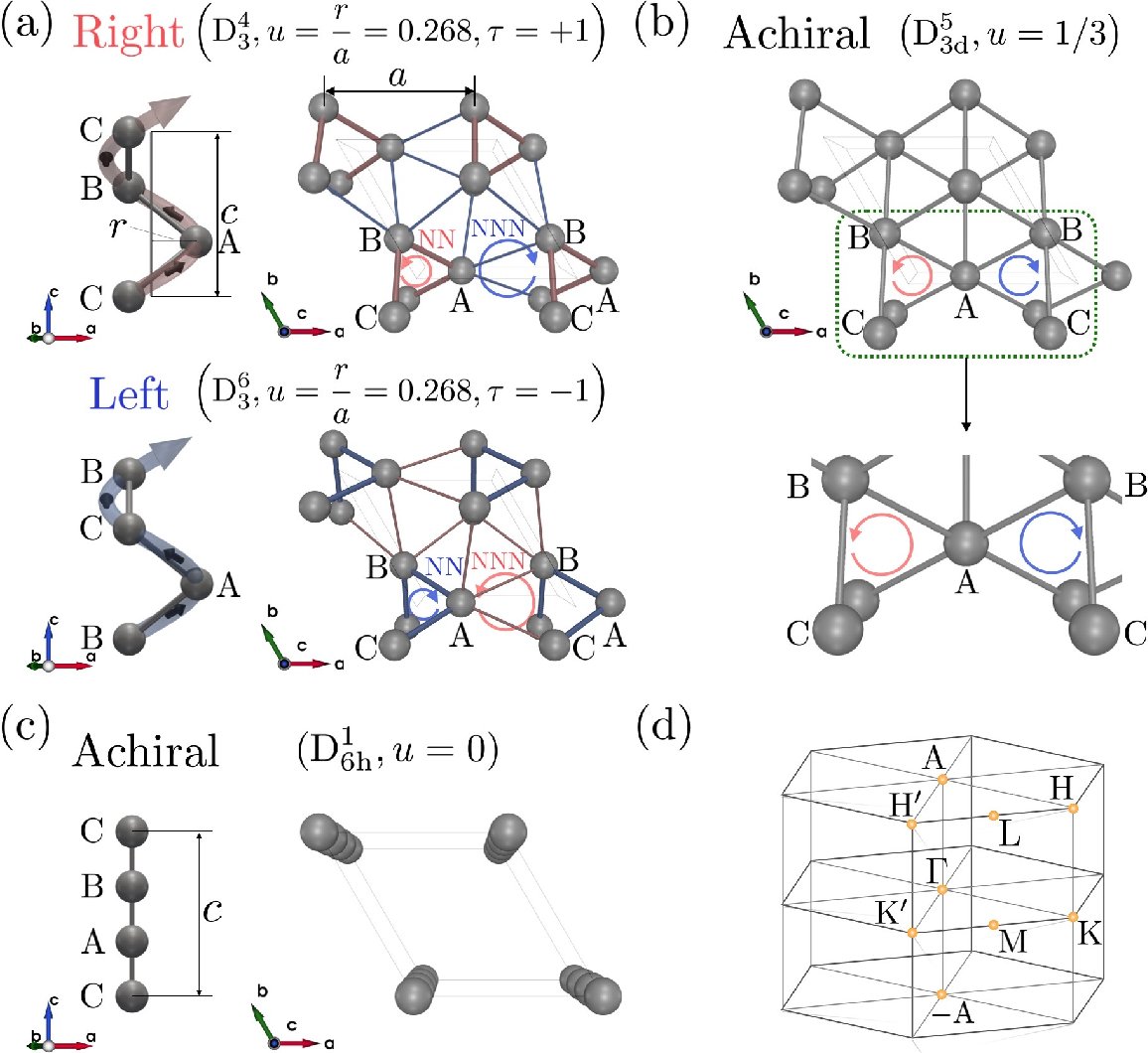}
  \vspace{3mm}
  \caption{
  \label{fig_crystal_structure}
  Crystal structures of Te and Se.
  (a) Equilibrium D$_{\rm 3}^{4}$ (right-handed) and D$_{\rm 3}^{6}$ (left-handed) chiral structures of Te.
   A unit cell contains three sublattices, $\bm{r}_{\rm A} = (u, 0, 0)$, $\bm{r}_{\rm B} = (0, u, +\tau/3)$, and $\bm{r}_{\rm C} = (-u, -u, -\tau/3)$ in the fractional coordinate, where $u = r/a$ is the dimensionless helix parameter and $r$ is the distance between the helical axis and each atoms, and $\tau = \pm 1$ corresponds to the right- and left-handed structures.
The red (blue) 
lines represent the nearest-neighbor (next-nearest-neighbor) and next-nearest-neighbor (nearest-neighbor) bonds in right- and left-handed structure, respectively.
   (b) Achiral D$_{\rm 3d}^{5}$ ($u = 1/3$) and (c) D$_{\rm 6h}^{1}$ ($u = 0$) structures.
  (d) First Brillouin zone of Te and Se.
  All figures for crystal structures and wave functions in this paper are drawn by QtDraw~\cite{HK_PRB_2023}.
 }
  \end{center}
\end{figure}

\begin{figure*}[t]
  \begin{center}
  \includegraphics[width=0.99\hsize]{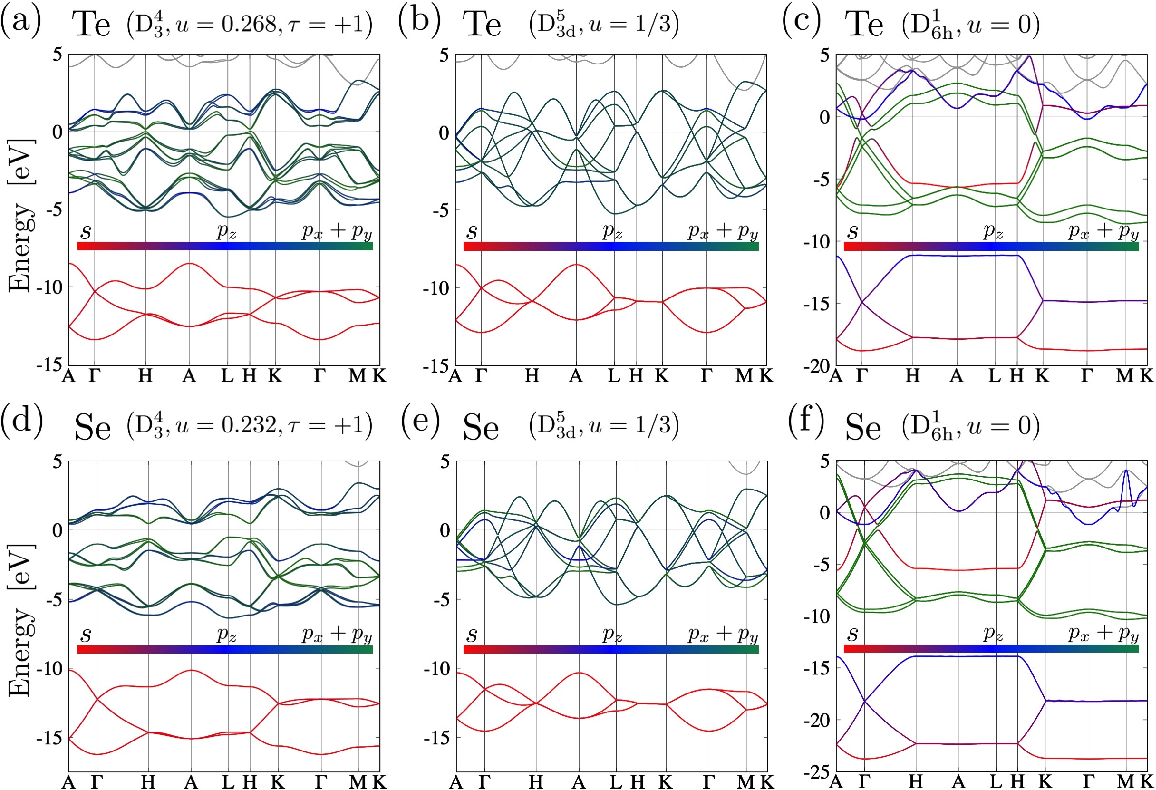}
  \vspace{6mm}
  \caption{
  \label{fig_band_structure}
  Band structures of Te and Se.
  The orbital dependence of the band dispersions obtained by CW models of (a)-(c) Te and (d)-(f) Se, where the red, blue, and green lines represents the CWFs with $s$, $p_{z}$, and $p_{x} + p_{y}$ orbital-like symmetry, as shown in Fig.~\ref{fig_cwfs}.
 The Fermi energy is set as the origin.
 }
   \end{center}
\end{figure*}

\begin{figure*}[t]
  \begin{center}
  \includegraphics[width=0.9\hsize]{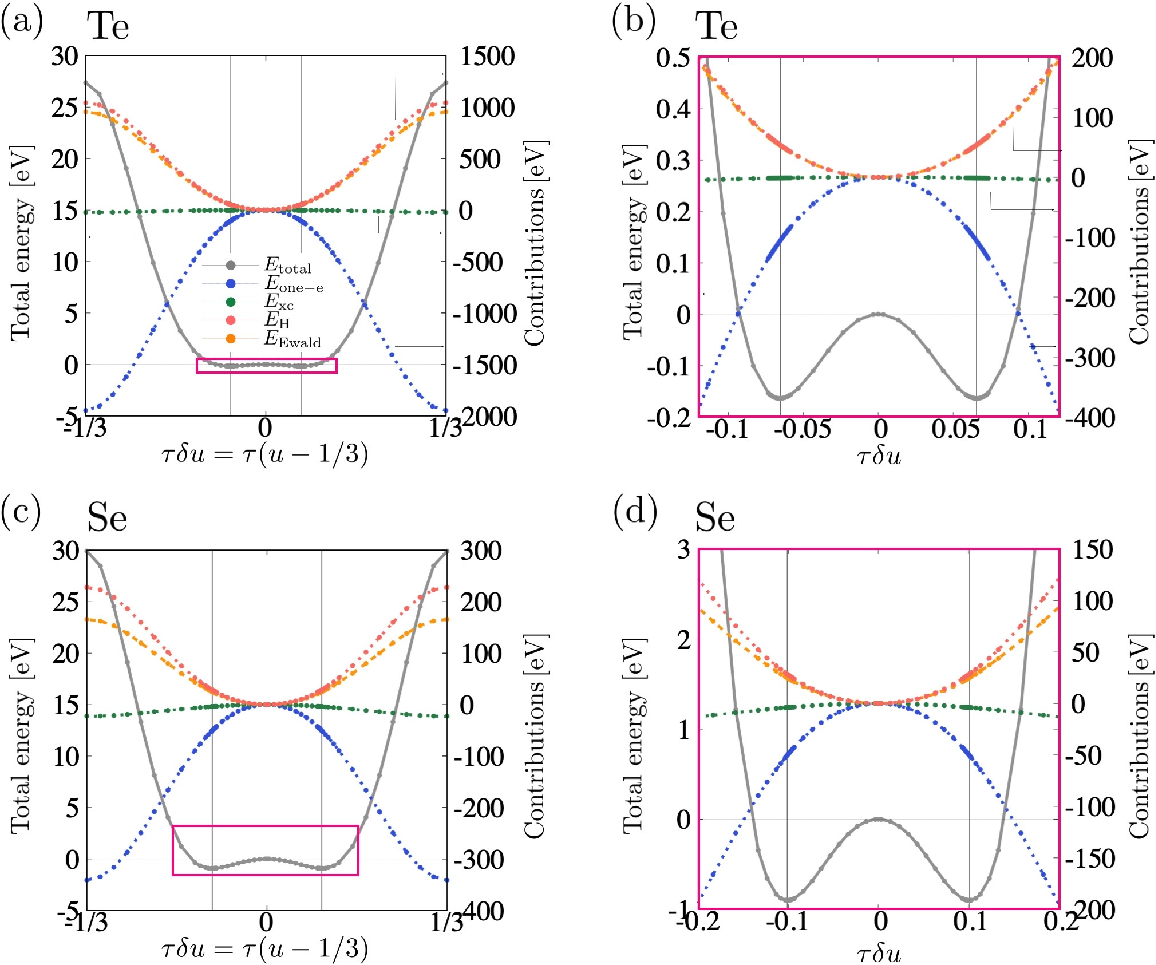}
  \vspace{3mm}
  \caption{
  \label{fig_Te_energy}
  $\tau \delta u$ ($\delta u = u - 1/3$) dependence of the total energy and its contributions (explained in detail in the main text) obtained by DFT calculations for (a), (b) Te and (c), (d) Se.
  (b) and (d) are the enlarged figures of the area surrounded by the pink solid line in (a) and (c), respectively.
  }
  \end{center}
\end{figure*}

In this section, we present an analysis of the stability mechanism of the helical structure of Te and Se based on the DFT calculations.
Figure~\ref{fig_crystal_structure}(a) shows the equilibrium right- and left-handed chiral structures of bulk Te or Se with the three-fold helical chains composed of A, B, and C sublattices in a unit cell.
The in-plane and out-of-plane lattice constants are slightly modified as $a = 4.6~\mathrm{\AA}$ and $c = 5.9~\mathrm{\AA}$ for Te in order to open finite band gap at H point~\cite{Nakazawa_PRM_2024} and $a = 4.3662~\mathrm{\AA}$ and $c = 4.9536~\mathrm{\AA}$ for Se~\cite{Teuchert_JPC_1975, Meijuan_PRB_2019}.
We introduce a dimensionless helix parameter $u = r/a$ $(0 \leq u \leq 1/3)$ where $r$ is the distance between the helical axis and each atom.
The atomic positions of the A, B, and C sublattices are $\bm{r}_{\rm A} = (u, 0, 0)$, $\bm{r}_{\rm B} = (0, u, +\tau/3)$, and $\bm{r}_{\rm C} = (-u, -u, -\tau/3)$ in the fractional coordinate, where $\tau = \pm 1$ corresponds to the right- and left-handed structures belonging to the space group P$3_{1}21$ ($\#$152, D$_{3}^{4}$) and P$3_{2}21$ ($\#$154, D$_{3}^{6}$).
When $0 < u < 1/3$ the crystal structure is chiral, and the equilibrium value of $u$ is $u = 0.268$ for Te and $u = 0.232$ for Se.
In the right-handed chiral structure, each atom is covalently bonded to the two intrachain nearest-neighbor (NN) atoms and is weekly bonded to the four interchain next-nearest-neighbor (NNN) atoms, as depicted by red and blue lines in the upper panel of Fig.~\ref{fig_crystal_structure}(a), while red and blue lines correspond to the NNN and NN bonds in the left-handed structure as shown in the lower panel of Fig.~\ref{fig_crystal_structure}(a).
As shown in Fig.~\ref{fig_crystal_structure}(b), when $u = 1/3$, the lengths of the NN and NNN bonds become the same, and the crystal structure belongs to achiral R-3m ($\#$166, D$_{\rm 3d}^{5}$), containing right- and left-handed chains of the same size.
On the other hand, when $u = 0$, the A, B, and C sublattices constitute one-dimensional straight chains with achiral structure in P6/mmm ($\#$191, D$_{\rm 6h}^{1}$) as shown in Fig.~\ref{fig_crystal_structure}(c).

We then perform 
fully relativistic self-consistent DFT calculations to obtain total energy of Te and Se by using Quantum ESPRESSO~\cite{Giannozzi_QE_2009}.
We use 
the PBE correlation functional~\cite{PBE_PRL_1996} and the full-relativistic optimized norm-conserving Vanderbilt (ONCV) pseudopotential~\cite{ONCV_PRB_2013} downloaded from PseudoDojo~\cite{PseudoDojo_2018}.
The kinetic energy cutoff of the KS orbitals and the convergence threshold are set to be 90 Ry and $1 \times 10^{-12}$ Ry, and the $\bm{k}$ grid is taken as $(N_{1}, N_{2}, N_{3}) = (12, 12, 16)$.

The DFT band structures of the D$_{3}^{4}$, D$_{\rm 3d}^{5}$, and D$_{\rm 6h}^{1}$ structures of Te and Se are shown in Fig.~\ref{fig_band_structure}(a)-(c) and (d)-(f), respectively.
The equilibrium Te and Se open finite band gap at H point and become semimetallic~\cite{Nakazawa_PRM_2024} in order to reduce the total energy of the system as shown in Fig.~\ref{fig_band_structure}(a) and (d).
On the other hand, the band structures of the achiral D$_{\rm 3d}^{5}$ structure are metallic and the electronic states at the Fermi level are degenerate as shown in Fig.~\ref{fig_band_structure}(b) and (e).
Here we introduce a displacement parameter $\delta u = u - 1/3$.
As shown in Figs.~\ref{fig_Te_energy}(a) and (b), the equilibrium right- or left-handed structure of Te with $\tau \delta u = \mp 0.06533$ is energetically most stable and close to the achiral D$_{\rm 3d}^{5}$ structure with $\delta u = 0$, whereas the achiral D$_{\rm 6h}^{1}$ structure with metallic band structure is energetically unstable.
These results are consistent with the previous studies~\cite{Ghosh_Helical_JPCC_2008, Honglie_PRB_2022}.
Se exhibits similar behavior, becoming energetically most stable at $\tau \delta u = \mp 0.1013$, as shown in Figs.~\ref{fig_Te_energy}(c) and (d).
Therefore, the equilibrium chiral structures of Te and Se can be regarded as arising from the Peierls distortion with respect to the achiral D$_{\rm 3d}^{5}$ structure with $\delta u = 0$~\cite{Tangney_PRB_2002}.
Indeed, the light-induced inverse Peierls distortion from equilibrium chiral structure to metastable achiral D$_{\rm 3d}^{5}$ one has been predicted theoretically~\cite{Kim_PRB_2003, Johnson_PRL_2009} and reported experimentally in Te using time-resolved optical second harmonic generation~\cite{Honglie_PRB_2022}.
Additionally, under high pressure, the achiral D$_{\rm 3d}^{5}$ structure of Te becomes energetically more stable than the chiral structure~\cite{Decker_ZfAAC_2002, Hejny_PRB_2006, Marini_PRB_2012}.

In order to identify which contribution in energy stabilizes the chiral structure of Te and Se, we decompose the total energy $E_{\rm total}$ into four contributions as follows:
\begin{align}
E_{\rm total} &= E_{\rm one-e} + E_{\rm xc} + E_{\rm H} + E_{\rm Ewald}.
\label{eq_E_total}
\end{align}
$E_{\rm one-e}$, $E_{\rm xc}$, $E_{\rm H}$, and $E_{\rm Ewald}$ are the one-electron contribution (electronic kinetic energy plus the pseudopotential energy), exchange-correlation energy, Hartree energy, and the Ewald summation of ion-ion Coulomb repulsion energy~\cite{Ewald_1921}, which are given in the atomic units ($e = \hbar = c = m_{e} = 1$) by using the Kohn-Sham (KS) orbital $\psi_{n\bm{k}}^{\rm KS}(\bm{r})$ and the electronic density $n(\bm{r}) = \sum_{n\bm{k}}^{\rm occupied} |\psi_{n\bm{k}}^{\rm KS}(\bm{r})|^{2}$ as
\begin{align}
&E_{\rm one-e} = \sum_{n\bm{k}} \int d\bm{r} \psi_{n\bm{k}}^{\rm KS *}(\bm{r}) \left(-\frac{\bm{\nabla}^{2}}{2}\right) \psi_{n\bm{k}}^{\rm KS}(\bm{r}) 
\cr & \qquad\qquad + \int d\bm{r} V_{\rm ps}(\bm{r}) n(\bm{r}),
\label{eq_E_one_e} \\
&E_{\rm xc} = \int d\bm{r} \epsilon_{\rm xc}(\bm{r}) n(\bm{r}),
\label{eq_E_xc} \\
&E_{\rm H} = \frac{1}{2} \int \int d\bm{r} d\bm{r}' \frac{n(\bm{r})n(\bm{r}')}{|\bm{r}-\bm{r}'|},
\label{eq_E_H} \\
&E_{\rm Ewald} = \frac{1}{2} \sum_{I \neq J}^{N_{n}} \frac{Z_{I} Z_{J}}{|\bm{R}_{I} - \bm{R}_{J}|},
\label{eq_E_Ewald}
\end{align}
where $V_{\rm ps}(\bm{r})$ represents the pseudopotential and $\epsilon_{\rm xc}(\bm{r})$ denotes the exchange-correlation potential.
$Z_{I}$ and $Z_{J}$ are the charge of the ionic core at $\bm{R}_{I}$ and $\bm{R}_{J}$, respectively.
As shown in Fig.~\ref{fig_Te_energy}, the dominant contribution in stabilizing the chiral structure of Te and Se is the one-electron contribution $E_{\rm one-e}$, and the exchange-correlation term $E_{\rm xc}$ gives a negligible contribution comparatively.
Meanwhile, the Hartree term $E_{\rm H}$ and the ionic contribution $E_{\rm Ewald}$ result in energy loss in the chiral structure.
Therefore, the chiral structure of Te and Se is stabilized mainly by the symmetry breaking of the electronic system that opens the one-body energy gap, as discussed above.
It is consistent with our analysis in Sec.~\ref{sec:results_discussions}, where the predominant order parameter for the chiral structure of Te and Se arises from one-body spin-independent hopping terms.

\section{Symmetry-adapted closest Wannier model}
\label{sec:scw}

\begin{figure}[t]
  \begin{center}
  \includegraphics[width=1.0\hsize]{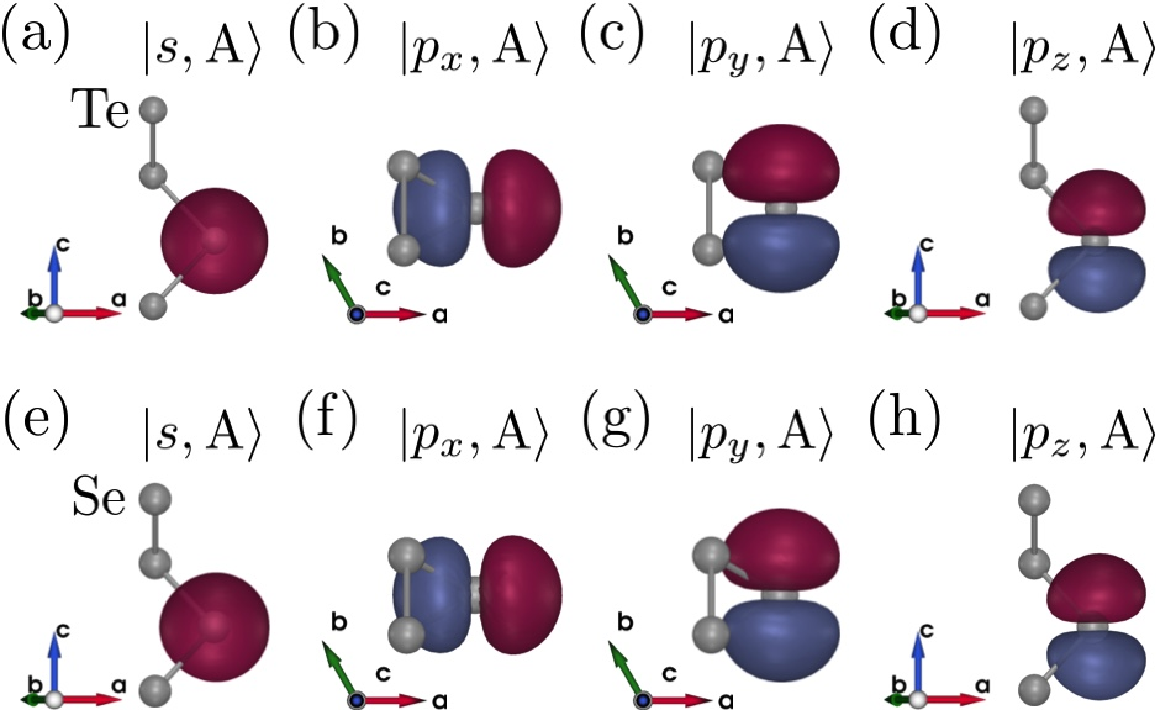}
  \vspace{2mm}
  \caption{
  \label{fig_cwfs}
  The CWFs with $s$, $p_{x}$, $p_{y}$, and $p_{z}$ like character localized at A site of (a)-(d) Te and (e)-(h) Se. 
  They are obtained for the equilibrium structures.
  }
  \end{center}
\end{figure}

\begin{figure*}[t]
  \begin{center}
  \includegraphics[width=0.9\hsize]{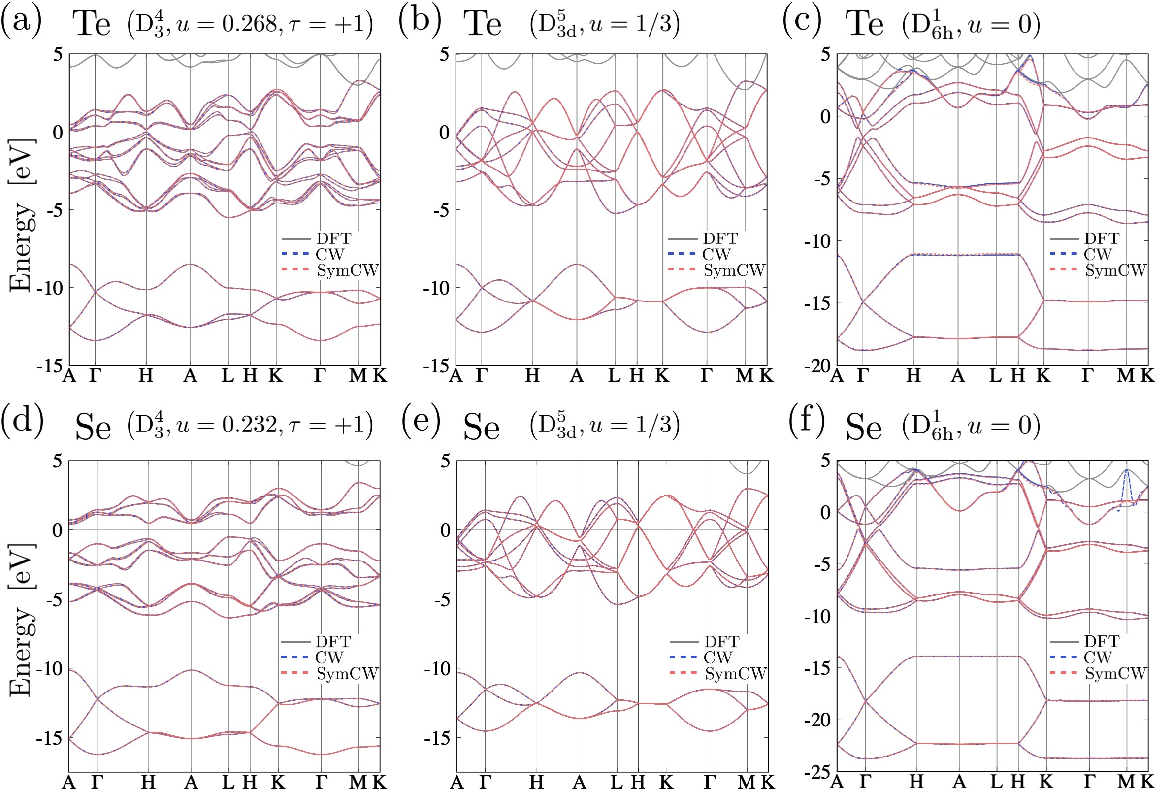}
  \vspace{2mm}
  \caption{
  \label{fig_scw_band}
The comparison of the band dispersions among the DFT calculation, CW 
 and SymCW models of (a)-(c) Te and (d)-(f) Se.
(a)-(c) and (d)-(f) correspond to Figs.~\ref{fig_band_structure}(a)-(c) and (d)-(f), respectively.
 }
  \end{center}
\end{figure*}

In this section, we construct the symmetry-adapted closest Wannier model for Te and Se.
Since the symmetry properties of the closest Wannier (CW) functions (CWFs) are common to those of the original atomic orbitals~\cite{Ozaki_CW_PRB_2024}, SAMBs can be defined as the complete orthonormal matrix basis set $\left\{\mathbb{Z}_{j}\right\}$ in the Hilbert space of the CWFs, $\mathrm{Tr}(\mathbb{Z}_{i}\mathbb{Z}_{j}) = \delta_{i,j}$.

The CW model is constructed by using the closest Wannier method~\cite{Ozaki_CW_PRB_2024} and our open-source Python library SymClosestWannier~\cite{RO_arxiv_2025}.
With regard to the initial guess of the CWFs, the hydrogenic atomic $s$ and three $p$ orbitals for each Te or Se atom are taken into account, as they primarily contribute to the electronic states near the Fermi level.
Figures~\ref{fig_cwfs}(a)-(d) and (e)-(h) represent the CWFs with $s$, $p_{x}$, $p_{y}$, and $p_{z}$ like character localized at A site of Te and Se, respectively.
Due to the closest property of the CWFs to the original atomic orbitals used for the initial guess, the obtained CWFs are well-localized at each atom.
Furthermore, it is confirmed that their symmetry properties are common to that of the original ones.

Utilizing the completeness and orthonormality of SAMBs, the $24\times 24$ ($1+3$ orbitals, 2 spins, and 3 sublattices) closest Wannier TB Hamiltonian matrix $H^{\rm CW}$ is rewritten as a linear combination of $\left\{\mathbb{Z}_{j}\right\}$ as 
\begin{align}
&H^{\rm SymCW} = \sum_{j}^{\Gamma_{j} \in {\rm A}_{1}} z_{j} \mathbb{Z}_{j},
\label{eq_SymCW}
\\
&z_{j} = \mathrm{Tr}\left[\mathbb{Z}_{j} H^{\rm CW}\right],
\label{eq_coeff}
\end{align}
where $\{\mathbb{Z}_{j}\}$ belong to the totally symmetric IR A$_{1}$ in D$_{3}$ point group, which are automatically generated by the Python library, MultiPie~\cite{HK_PRB_2023}.
We refer to this symmetrized model as the symmetry-adapted closest Wannier (SymCW) model.
As shown later, $\mathbb{Z}_{j}$ is expressed by the linear combination of the direct product of the atomic multipole bases $X^{\rm (a)}$ and the site- or bond-cluster multipole bases $Y^{\rm (s/b)}$~\cite{HK_PRB_2023}.
In this study, by considering up to the 258 bond-clusters (the maximum bond length is about the distance between the home unit-cell and the 4th neighbor unit-cell) and 15,252 independent SAMBs, the mean absolute error between the energy eigenvalues obtained from the SymCW model and the DFT calculation for Te and Se are
\begin{align}
L(\bm{z}) = \sum_{\bm{k}} \sum_{n=1}^{N_{\rm B}}  \frac{|\epsilon_{n\bm{k}}^{\rm DFT} - \epsilon_{n\bm{k}}^{\rm SymCW}|}{N_{k}N_{\rm B}}  = 
\begin{cases}
1.07\, [\mathrm{meV}]  \quad\,\,\, (\mathrm{Te}), \\
0.258\, [\mathrm{meV}] \quad (\mathrm{Se}),
\end{cases}
\end{align}
where $\epsilon_{n\bm{k}}^{\rm DFT}$ and $\epsilon_{n\bm{k}}^{\rm SymCW}$ are the $n$-th energy eigenvalue of the DFT band dispersion and our SymCW model.
$N_{k} = 2304\, (12\times 12 \times 16)$ and $N_{\rm B} =24$ represent the number of $\bm{k}$ points and the number of the eigenvalues, respectively.
The comparison of the energy dispersions is shown in Fig.~\ref{fig_scw_band}, where the CW and SymCW models well reproduce the DFT band dispersion.
Note that in the previous study~\cite{RO_PRL_2022}, only 255 independent parameters were used in Eq.~(\ref{eq_SymCW}), which was not enough to accurately reproduce the DFT bands.

The linear coefficients of each SAMB $\left\{z_{j}\right\}$ (model parameters) are determined through simple matrix projection given by Eq.~(\ref{eq_coeff}) without any iterative procedure.
They are interpreted as the crystalline electric fields (CEF), SOC, and electron hoppings, and so on.
In this way, the present method enables us to quantitatively identify the dominant electronic multipole basis in $H^{\rm SymCW}$. 
In particular, we can quantitatively identify the predominant order parameter for structural chirality by evaluating the expectation value of the ET monopoles and quadrupoles, $\braket{\mathbb{Z}_{j}}$, as discussed in the following sections.

\section{Results and discussions}
\label{sec:results_discussions}

In this section, we present our analysis on the order parameters for structural chirality of Te and Se.
To identify the  predominant order parameter, we compute the evolution of order parameters with changing helix parameter $\tau \delta u$ based on the SymCW model~\cite{RO_arxiv_2025}.
Our main results are (i) both localized and itinerant spinless ET quadrupoles are proper indicators of structural chirality, since they are non-zero only when the system has chiral structure, and its sign depends on the handedness of chiral structure, and (ii) the itinerant spinless ET quadrupole is the predominant order parameter because its expectation value is largest among chiral order parameters. 
Moreover, we elucidate that the monopole-like OAM texture in the momentum space arises from the itinerant spinless ET quadrupole without SOC.

\subsection{Physical interpretation of SAMBs}
\label{sec:samb}

\begin{figure*}[t]
  \begin{center}
  \includegraphics[width=1\hsize]{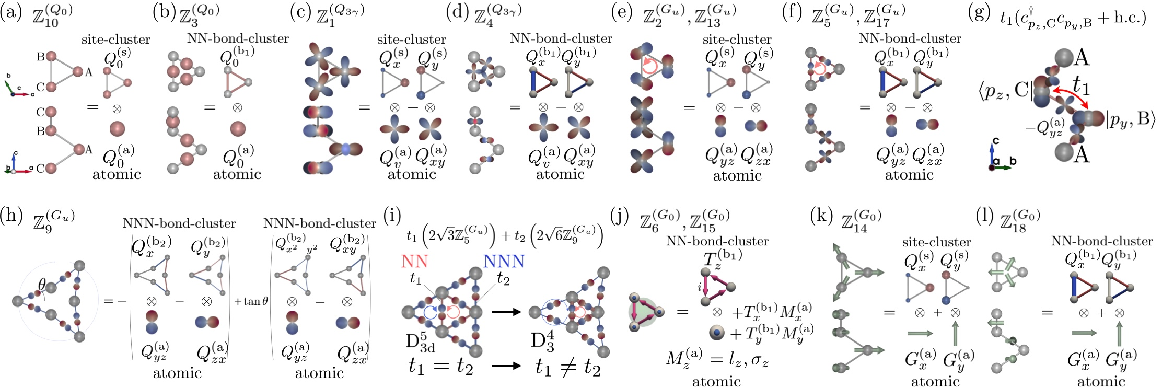}
  \vspace{3mm}
  \caption{
  \label{fig_samb_def}
 Schematic representations  of SAMBs. 
 (a) Site-cluster and (b) NN bond-cluster E monopoles, $\mathbb{Z}_{10}^{(Q_{0})}$ and $\mathbb{Z}_{3}^{(Q_{0})}$.
 (c) Site-cluster and (d) NN bond-cluster E octupoles, $\mathbb{Z}_{1}^{(Q_{3\gamma})}$ and $\mathbb{Z}_{4}^{(Q_{3\gamma})}$.
 (e) Site-cluster ET quadrupoles, $\mathbb{Z}_{2}^{(G_{u})}$ and $\mathbb{Z}_{13}^{(G_{u})}$,  and (f) NN bond-cluster ET quadrupoles, $\mathbb{Z}_{5}^{(G_{u})}$ and $\mathbb{Z}_{17}^{(G_{u})}$.
 (g) Spin-independent off-diagonal real hopping between $p_{z}$ orbital at C site and $p_{y}$ orbital at B site, which is induced by $\mathbb{Z}_{5}^{(G_{u})}$.
 (h) NNN bond-cluster ET quadrupole, $\mathbb{Z}_{9}^{(G_{u})}$, where $\theta$ denotes the angle between the $x$-axis and the NNN bond between the B site in the home unit-cell and the A site in the NN unit-cell with $\bm{R} = -\bm{a}_{1}$.
 (i) When the system is achiral D$_{\rm 3d}$ structure, the NN and NNN hoppings $t_{1}$ and $t_{2}$ become the same, whereas $t_{1} \neq t_{2}$ when the system is chiral structures.
 (j) NN bond-cluster ET monopoles, $\mathbb{Z}_{6}^{(G_{0})}$ and $\mathbb{Z}_{15}^{(G_{0})}$.
 (k) Site-cluster ET monopole, $\mathbb{Z}_{14}^{(G_{0})}$ and (l) bond-cluster ET monopole, $\mathbb{Z}_{18}^{(G_{0})}$.
  }
  \end{center}
\end{figure*}

Let us present the physical interpretation of each SAMB and the computational results of the equilibrium right-handed structure ($\tau = +1$) of Te with $u = 0.268$ and Se with $u = 0.232$.
Equation~(\ref{eq_SymCW}) is decomposed as 
\begin{align}
H^{\rm SymCW} &= H_{\rm CEF} + H_{\rm SOC} + \sum_{n = 1}^{258} \left(H_{\rm hop}^{\rm (b_{n})} + H_{\rm hop}^{\rm (b_{n})'}\right),
\label{eq_SymCW_2}
\end{align}
where $H_{\rm CEF}$ and $H_{\rm SOC}$ are the CEF and SOC terms, while $H_{\rm hop}^{\rm (b_{n})}$ and $H_{\rm hop}^{\rm (b_{n})'}$ are the spin-independent and spin-dependent hopping terms in the $n$-th bond-cluster.
There are 52 parameters within the NN hopping: 6 CEF parameters, 10 SOC parameters, and 36 NN intrachain hopping parameters.
The spinless terms are given by
\begin{align}
H_{\rm CEF} &= \Delta_{1}^{(Q_{3\gamma})} \mathbb{Z}_{1}^{(Q_{3\gamma})} + \Delta_{2}^{(G_{u})} \mathbb{Z}_{2}^{(G_{u})} + \cdots,
\\
H_{\rm hop}^{\rm (b_{1})} &= t_{3}^{(Q_{0})} \mathbb{Z}_{3}^{(Q_{0})} +  t_{4}^{(Q_{3\gamma})} \mathbb{Z}_{4}^{(Q_{3\gamma})} + t_{5}^{(G_{u})} \mathbb{Z}_{5}^{(G_{u})}
\cr & \quad + t_{6}^{(G_{0})} \mathbb{Z}_{6}^{(G_{0})} + t_{7}^{(G_{u})} \mathbb{Z}_{7}^{(G_{u})}  + \cdots,
\\
H_{\rm hop}^{\rm (b_{2})} &= t_{8}^{(Q_{0})} \mathbb{Z}_{8}^{(Q_{0})} + t_{9}^{(G_{u})} \mathbb{Z}_{9}^{(G_{u})}  + \cdots,
\end{align}
and the spinful terms are given by
\begin{align}
H_{\rm SOC} &= \xi_{10}^{(Q_{0})} \mathbb{Z}_{10}^{(Q_{0})} + \xi_{11}^{(G_{0})} \mathbb{Z}_{11}^{(G_{0})} + \xi_{12}^{(G_{u})} \mathbb{Z}_{12}^{(G_{u})} 
\cr & \quad + \xi_{13}^{(G_{u})} \mathbb{Z}_{13}^{(G_{u})} + \xi_{14}^{(G_{0})} \mathbb{Z}_{14}^{(G_{0})} + \cdots,
\\
H_{\rm hop}^{\rm (b_{1})'} &= \alpha_{15}^{(G_{0})} \mathbb{Z}_{15}^{(G_{0})} + \alpha_{16}^{(G_{u})} \mathbb{Z}_{16}^{(G_{u})} + \alpha_{17}^{(G_{u})} \mathbb{Z}_{17}^{(G_{u})}
\cr & \quad + \alpha_{18}^{(G_{0})} \mathbb{Z}_{18}^{(G_{0})} + \cdots.
\end{align}
Among them, we focus on the significant contributions including the ET monopole $G_{0}$ and quadrupole $G_{u}$ SAMBs.
The explicit definitions of each SAMB are summarized in Table~\ref{tab_SAMB_Te}, and their schematic picture is shown in Fig.~\ref{fig_samb_def}.

In Table~\ref{tab_SAMB_Te}, the spinless atomic E monopole and quadrupoles, $Q_{0,p}^{\rm (a)}$, $Q_{v}^{\rm (a)}$ ($v = x^{2}-y^{2}$), $Q_{xy}^{\rm (a)}$, $Q_{yz}^{\rm (a)}$, $Q_{zx}^{\rm (a)}$ and the dimensionless orbital angular momentum $\bm{l}$ in the $(p_{x},p_{y},p_{z})$ orbitals are defined as
\begin{align}
&Q_{v}^{\rm (a)} = \frac{1}{2} \begin{pmatrix} 1 & 0 & 0 \\ 0 & -1 & 0 \\ 0 & 0 & 0 \end{pmatrix} \otimes \sigma_{0},
\,
Q_{xy}^{\rm (a)} = \frac{1}{2} \begin{pmatrix} 0 & 1 & 0 \\ 1 & 0 & 0 \\ 0 & 0 & 0 \end{pmatrix} \otimes \sigma_{0},
\\
&Q_{yz}^{\rm (a)} = \frac{1}{2} \begin{pmatrix} 0 & 0 & 0 \\ 0 & 0 & 1 \\ 0 & 1 & 0 \end{pmatrix} \otimes \sigma_{0},
\,
Q_{zx}^{\rm (a)} = \frac{1}{2} \begin{pmatrix} 0 & 0 & 1 \\ 0 & 0 & 0 \\ 1 & 0 & 0 \end{pmatrix} \otimes \sigma_{0},
\\
&Q_{0,p}^{\rm (a)} =  \frac{1}{\sqrt{6}} \begin{pmatrix} 1 & 0 & 0 \\ 0 & 1 & 0 \\ 0 & 0 & 1 \end{pmatrix} \otimes \sigma_{0},
\,
l_{z} = \frac{1}{2} \begin{pmatrix} 0 & -i & 0 \\ i & 0 & 0 \\ 0 & 0 & 0 \end{pmatrix} \otimes \sigma_{0},
\\
&l_{x} =   \frac{1}{2} \begin{pmatrix} 0 & 0 & 0 \\ 0 & 0 & -i \\ 0 & i & 0 \end{pmatrix} \otimes \sigma_{0},
\,
l_{y} = \frac{1}{2} \begin{pmatrix} 0 & 0 & i \\ 0 & 0 & 0 \\ -i & 0 & 0 \end{pmatrix} \otimes \sigma_{0},
\end{align}
where $\sigma_{0}$ is the $2\times 2$ identity matrix in the spin space.
The spinless MT dipole $\bm{t} = \bm{r} \times \bm{l}$ in the $s$-$p$ Hybrid orbitals space $\braket{s|p}$ is defined as
\begin{align}
&t_{x} = \frac{1}{\sqrt{2}}  \begin{pmatrix} i & 0 & 0 \end{pmatrix} \otimes \sigma_{0},
\label{eq_tx_atomic}
\\
&t_{y} = \frac{1}{\sqrt{2}}  \begin{pmatrix} 0 & i & 0 \end{pmatrix} \otimes \sigma_{0},
\label{eq_ty_atomic}
\\
&t_{z} = \frac{1}{\sqrt{2}}  \begin{pmatrix} 0 & 0 & i \end{pmatrix} \otimes \sigma_{0}.
\label{eq_tz_atomic}
\end{align}
Furthermore, the spinful atomic multipole bases, $Q_{0}^{\rm (a)'}$, $G_{0}^{\rm (a)'}$, $G_{u}^{\rm (a)'}$, $G_{u}^{\rm (a)'}$, $\bm{G}^{\rm (a)'}$, and $Q_{\mu\nu}^{\rm (a)'}$ ($\mu, \nu = x,y,z$) are defined by
\begin{align}
&Q_{0}^{\rm (a)'} = \frac{1}{\sqrt{3}} \bm{l} \cdot \bm{\sigma},
\label{eq_Q0_atomic}
\\
&Q_{\mu\nu}^{\rm (a)'} = \frac{1}{\sqrt{2}} \left( l_{\mu} \sigma_{\nu} + l_{\nu} \sigma_{\mu} \right),
\label{eq_Q2_sf_atomic}
\\
&\bm{G}^{\rm (a)'} = \frac{1}{\sqrt{2}} \bm{l}\times \bm{\sigma}, 
\label{eq_G_atomic}
\\
&G_{0}^{\rm (a)'} = \frac{1}{\sqrt{3}} \bm{t} \cdot \bm{\sigma} = \frac{(\bm{r} \times \bm{l}) \cdot \bm{\sigma} }{\sqrt{3}} = \frac{\bm{r} \cdot (\bm{l} \times \bm{\sigma})}{\sqrt{3}} =  \frac{1}{\sqrt{3}} \bm{r} \cdot \bm{G}^{\rm (a)'}, 
\label{eq_G0_atomic}
\\
&G_{u}^{\rm (a)'} = \frac{1}{\sqrt{6}} \left(3 t_{z} \sigma_{z} - \bm{t} \cdot \bm{\sigma} \right).
\label{eq_Gu_atomic}
\end{align}
Here, the spinful E monopole $Q_{0}^{\rm (a)'} \propto \bm{l} \cdot \bm{\sigma}$ corresponds to the atomic SOC, while $Q_{\mu\nu}^{\rm (a)'}$ and $\bm{G}^{\rm (a)'}$ are the spinful E quadrupole and ET dipole consisting of the symmetric and antisymmetric product of the orbital and spin angular momenta, respectively.
The ET monopole $G_{0}^{\rm (a)'}$ is defined both in the forms of $\bm{T}\cdot\bm{M}$ and $\bm{Q}\cdot\bm{G}$, and they are depicted already in Fig.~\ref{fig_ET}(b) and (a). 
The ET quadrupole $G_{u}^{\rm (a)'}$ represents monoaxial anisotropy along $z$ axis.
Note that $G_{0}^{\rm (a)'}$ is active in all chiral point groups, while $G_{u}^{\rm (a)'}$ is active only in monoaxial chiral point groups.

$Q_{0}^{\rm (s)}$ and $Q_{\mu}^{\rm (s)}$ ($\mu = x, y$) are the site-cluster E monopole and $\mu$ component of the E dipole, and $Q_{\mu}^{\rm (b_{1})}$ ($\mu = x, y$) and $T_{\mu}^{\rm (b_{1})}$ ($\mu = x, y, z$) are the $\mu$ component of the NN bond-cluster E and MT dipoles, which are defined in the ABC sublattice space as
\begin{align}
&Q_{0}^{\rm (s)}=\frac{1}{\sqrt{3}} \begin{pmatrix} 1 & 0 & 0 \\ 0 & 1 & 0 \\ 0 & 0 & 1 \end{pmatrix},
\\
&Q_{x}^{\rm (s)}=\frac{1}{\sqrt{6}} \begin{pmatrix} 2 & 0 & 0 \\ 0 & -1 & 0 \\ 0 & 0 & -1 \end{pmatrix},
\, Q_{y}^{\rm (s)}=\frac{1}{\sqrt{2}} \begin{pmatrix} 0 & 0 & 0 \\ 0 & 1 & 0 \\ 0 & 0 & -1 \end{pmatrix},
\\
&Q_{x}^{\rm (b_{1})} = \frac{1}{2\sqrt{3}} \begin{pmatrix} 0 & 1 & 1 \\ 1 & 0 & -2 \\ 1 & -2 & 0 \end{pmatrix},
\, Q_{y}^{\rm (b_{1})} = \frac{1}{2} \begin{pmatrix} 0 & 1 & -1 \\ 1 & 0 & 0 \\ -1 & 0 & 0 \end{pmatrix},
\\
&Q_{0}^{\rm (b_{1})} = \frac{1}{\sqrt{6}} \begin{pmatrix} 0 & 1 & 1 \\ 1 & 0 & 1 \\ 1 & 1 & 0 \end{pmatrix},
\, T_{z}^{\rm (b_{1})} = \frac{1}{\sqrt{6}} \begin{pmatrix} 0 & -i & i \\ i & 0 & -i \\ -i & i & 0 \end{pmatrix},
\\
&T_{x}^{\rm (b_{1})} = \frac{1}{2} \begin{pmatrix} 0 & i & i \\ -i & 0 & 0 \\ -i & 0 & 0 \end{pmatrix},
\, T_{y}^{\rm (b_{1})} = \frac{1}{2\sqrt{3}} \begin{pmatrix} 0 & -i & i \\ i & 0 & 2i \\ -i & -2i & 0 \end{pmatrix}.
\end{align}

\begin{table*}[t]
\begin{center}
\caption{ \label{tab_SAMB_Te}
SAMBs and their coefficients of equilibrium Te and Se in $H_{\rm CEF}$, $H_{\rm SOC}$, $H_{\rm hop}^{\rm (b_{1})}$, $H_{\rm hop}^{\rm (b_{2})}$, and $H_{\rm hop}^{\rm (b_{1})'}$.
The symbols $X^{\rm (a)}$ and $Y^{\rm (s/b)}$ represent the atomic multipole and site- or bond-cluster multipole bases~\cite{HK_PRB_2023}, respectively.
$X^{\rm (a)}$ is defined in the $\braket{s|s}$, $\braket{s|p}$, and $\braket{p|p}$ orbitals, where the prime denotes the spinful term.
The upper and lower parts separated by the double line represent the spinless and spinful SAMBs, respectively.
The angle $\theta$ is defined in Fig.~\ref{fig_samb_def}(h).
}
\begin{ruledtabular}
\begin{tabular}{llcccc}
 & SAMB  & ($\mathcal{H}$, Cluster) & Coefficients & Weight in Te [eV] & Weight in Se [eV] \\
\hline
$H_{\rm CEF}$ & $\mathbb{Z}_{1}^{(Q_{3\gamma})} =  \frac{1}{\sqrt{2}} \left(Q_{x}^{\rm (s)} \otimes Q_{v}^{\rm (a)} - Q_{y}^{\rm (s)} \otimes Q_{xy}^{\rm (a)} \right)$ & ($\braket{p|p}, {\rm C}$) & $\Delta_{1}^{(Q_{3\gamma})}$  & $5.71\times 10^{-2}$ & $-6.36\times 10^{-2}$
\\
& $\mathbb{Z}_{2}^{(G_{u})} = \frac{1}{\sqrt{2}} \left(Q_{x}^{\rm (s)} \otimes Q_{yz}^{\rm (a)} - Q_{y}^{\rm (s)} \otimes Q_{zx}^{\rm (a)} \right)$ & ($\braket{p|p}, {\rm C}$) & $\Delta_{2}^{(G_{u})}$ & $1.49\times 10^{-1}$ & $2.04\times 10^{-1}$
\\
\hline
$H_{\rm hop}^{\rm (b_{1})}$ & $\mathbb{Z}_{3}^{(Q_{0})} = Q_{0}^{\rm (b_{1})} \otimes Q_{0,p}^{\rm (a)}$ & ($\braket{p|p}, {\rm b_{1}}$) & $t_{3}^{\rm (Q_{0})}$ & $1.93$ & $2.13$
\\
& $\mathbb{Z}_{4}^{(Q_{3\gamma})} = \frac{1}{\sqrt{2}} \left(Q_{x}^{\rm (b_{1})} \otimes Q_{v}^{\rm (a)} - Q_{y}^{\rm (b_{1})} \otimes Q_{xy}^{\rm (a)} \right)$ & ($\braket{p|p}, {\rm b_{1}}$) &  $t_{4}^{\rm (Q_{3\gamma})}$ & $3.63$ & $4.88$
\\
& $\mathbb{Z}_{5}^{(G_{u})} = \frac{1}{\sqrt{2}} \left(Q_{x}^{\rm (b_{1})} \otimes Q_{yz}^{\rm (a)} - Q_{y}^{\rm (b_{1})} \otimes Q_{zx}^{\rm (a)} \right)$ & ($\braket{p|p}, {\rm b_{1}}$) & $t_{5}^{(G_{u})}$ & $6.76$ & $9.35$
\\
& $\mathbb{Z}_{6}^{(G_{0})} =  \frac{1}{\sqrt{3}} \bm{T}^{\rm (b_{1})} \cdot \bm{l}$ & ($\braket{p|p}, {\rm b_{1}}$) & $t_{6}^{(G_{0})}$ & $3.43 \times 10^{-3}$ & $4.53\times 10^{-2}$
\\
& $\mathbb{Z}_{7}^{(G_{u})} = \frac{1}{\sqrt{6}} \left( 3 T_{z}^{\rm (b_{1})} l_{z} - \bm{T}^{\rm (b_{1})} \cdot \bm{l} \right)$ & ($\braket{p|p}, {\rm b_{1}}$) & $t_{7}^{(G_{u})}$ & $-3.69 \times 10^{-2}$ & $-5.03 \times 10^{-2}$
\\
\hline
$H_{\rm hop}^{\rm (b_{2})}$ & $\mathbb{Z}_{8}^{(Q_{0})} = \frac{1}{\sqrt{2}}
\left[
\sin\theta \left( Q_{x}^{\rm (b_{2})} \otimes Q_{yz}^{\rm (a)} - Q_{y}^{\rm (b_{2})} \otimes Q_{zx}^{\rm (a)} \right) \right.$ & ($\braket{p|p}, {\rm b_{2}}$) & $t_{8}^{(Q_{0})}$ & $-1.89\times 10^{-4}$ & $-5.43\times 10^{-2}$\\
& $\qquad\qquad \left. + \cos\theta\left( Q_{x^{2}-y^{2}}^{\rm (b_{2})} \otimes Q_{yz}^{\rm (a)} - Q_{xy}^{\rm (b_{2})} \otimes Q_{zx}^{\rm (a)} \right)\right]$
\\
& $\mathbb{Z}_{9}^{(G_{u})} = -\frac{1}{\sqrt{2}} 
\left[
\cos\theta\left( Q_{x}^{\rm (b_{2})} \otimes Q_{yz}^{\rm (a)} - Q_{y}^{\rm (b_{2})} \otimes Q_{zx}^{\rm (a)} \right) \right.$ & ($\braket{p|p}, {\rm b_{2}}$) & $t_{9}^{(G_{u})}$ & $4.10$ & $3.11$
\\
& $\qquad\qquad \left. -\sin\theta
 \left( Q_{x^{2}-y^{2}}^{\rm (b_{2})} \otimes Q_{yz}^{\rm (a)} - Q_{xy}^{\rm (b_{2})} \otimes Q_{zx}^{\rm (a)} \right)\right]$
\\
\hline \hline
$H_{\rm SOC}$ & $\mathbb{Z}_{10}^{(Q_{0})} = Q_{0}^{\rm (s)} \otimes Q_{0}^{\rm (a)'}$ & ($\braket{p|p}, {\rm C}$) & $\xi_{10}^{(Q_{0}})$ & $1.63$ & $0.765$
\\
& $\mathbb{Z}_{11}^{(G_{0})} = Q_{0}^{\rm (s)} \otimes G_{0}^{\rm (a)'}$ & ($\braket{s|p}, {\rm C}$) & $\xi_{11}^{(G_{0}})$  & $-4.53\times 10^{-4}$ & $-1.59\times 10^{-4}$
\\
& $\mathbb{Z}_{12}^{(G_{u})} = Q_{0}^{\rm (s)} \otimes G_{u}^{\rm (a)'}$ & ($\braket{s|p}, {\rm C}$) & $\xi_{12}^{(G_{u}})$ & $-4.98\times 10^{-3}$ & $1.06\times 10^{-3}$
\\
& $\mathbb{Z}_{13}^{(G_{u})} = \frac{1}{\sqrt{2}} \left(Q_{x}^{\rm (s)} \otimes Q_{yz}^{\rm (a)'} - Q_{y}^{\rm (s)} \otimes Q_{zx}^{\rm (a)'} \right)$ & ($\braket{p|p}, {\rm C}$) & $\xi_{13}^{(G_{u}})$ & $8.42\times 10^{-3}$ & $7.05\times 10^{-3}$
\\
& $\mathbb{Z}_{14}^{(G_{0})} = \frac{1}{\sqrt{2}} \left(Q_{x}^{\rm (s)} \otimes G_{x}^{\rm (a)'} + Q_{y}^{\rm (s)} \otimes G_{y}^{\rm (a)'} \right)$ & ($\braket{p|p}, {\rm C}$) & $\xi_{14}^{(\rm G_{0}})$ & $-3.33\times 10^{-3}$  & $1.87\times 10^{-4}$
\\
\hline
$H_{\rm hop}^{\rm (b_{1})'}$ & $\mathbb{Z}_{15}^{(G_{0})} = \frac{1}{\sqrt{3}} \bm{T}^{\rm (b_{1})} \cdot \bm{\sigma}$ & ($\braket{p|p}, {\rm b_{1}}$) & $\alpha_{15}^{(G_{0})}$ & $-1.15 \times 10^{-4}$ & $4.57 \times 10^{-4}$
\\
& $\mathbb{Z}_{16}^{(G_{u})} =  \frac{1}{\sqrt{6}} \left( 3 T_{z}^{\rm (b_{1})} \sigma_{z} - \bm{T}^{\rm (b_{1})} \cdot \bm{\sigma} \right)$ & ($\braket{p|p}, {\rm b_{1}}$) & $\alpha_{16}^{(G_{u})}$ & $4.92 \times 10^{-6}$ & $2.28\times 10^{-3}$
\\
& $\mathbb{Z}_{17}^{(G_{u})} = \frac{1}{\sqrt{2}} \left(Q_{x}^{\rm (b_{1})} \otimes Q_{yz}^{\rm (a)'} - Q_{y}^{\rm (b_{1})} \otimes Q_{zx}^{\rm (a)'} \right)$ & ($\braket{p|p}, {\rm b_{1}}$) & $\alpha_{17}^{(G_{u})}$ & $7.02 \times 10^{-2}$ & $1.43\times 10^{-2}$
\\
& $\mathbb{Z}_{18}^{(G_{0})} = \frac{1}{\sqrt{2}} \left(Q_{x}^{\rm (b_{1})} \otimes G_{x}^{\rm (a)'} + Q_{y}^{\rm (b_{1})} \otimes G_{y}^{\rm (a)'} \right)$ & ($\braket{p|p}, {\rm b_{1}}$) & $\alpha_{18}^{(G_{0})}$& $-3.48 \times 10^{-3}$ & $2.03\times 10^{-3}$
\end{tabular}
\end{ruledtabular}
\end{center}
\end{table*}

\begin{figure*}[t]
  \begin{center}
  \includegraphics[width=0.99\hsize]{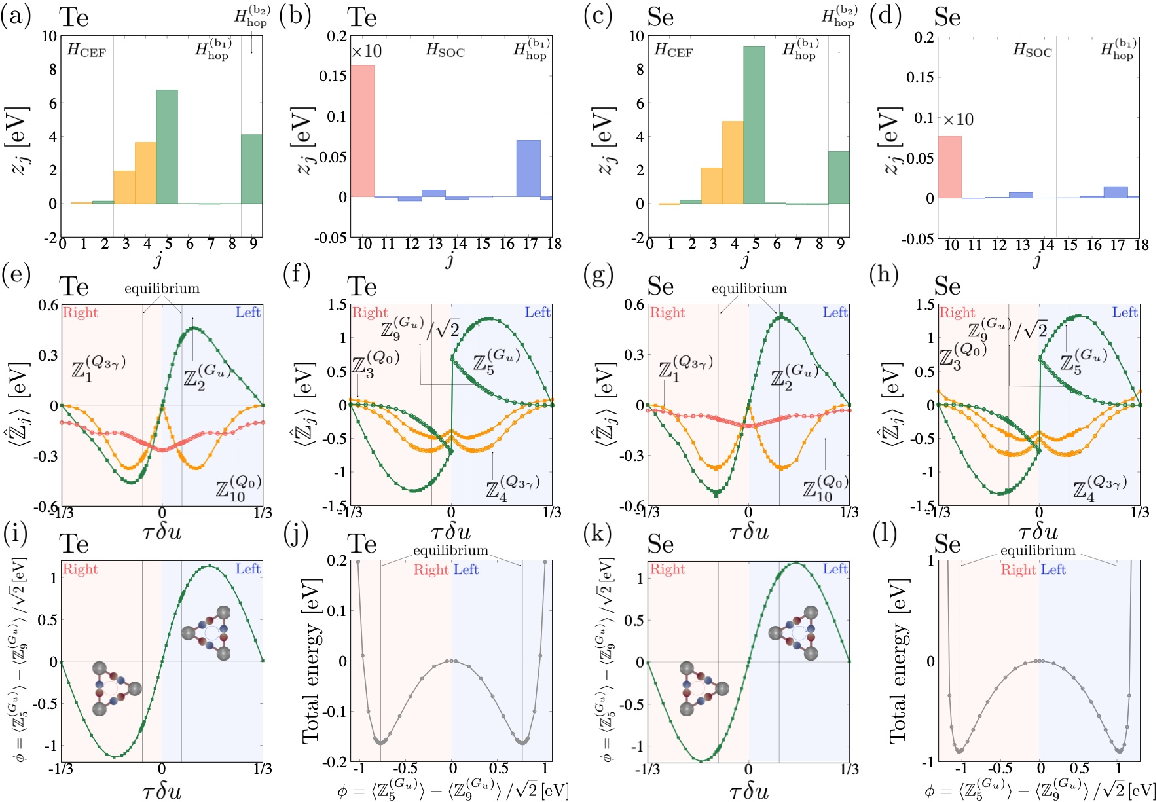}
  \hspace{3mm}
  \caption{
  \label{fig_samb_coeff}
  The coefficients of (a) spinless and (b) spinful SAMBs of the equilibrium Te, and (c), (d) Se. 
  The displacement $\tau \delta u$ dependence of the expectation value of the (e) localized and (f) itinerant SAMBs for the ground state of Te, and (g), (h) Se.
  The orange, green, red, and blue bars and lines in (a)-(h) correspond to the spinless E, spinless ET, spinful E, and spinful ET multipoles, respectively.
  The $\tau \delta u$ dependence of the expectation value of the net itinerant ET quadrupole $\phi$ of (i) Te and (k) Se.
  The variation of the total energy of (j) Te and (l) Se, obtained by DFT calculation [See Eq.~(\ref{eq_E_total}) and Fig.~\ref{fig_Te_energy}], as a function of $\phi = \braket{\mathbb{Z}_{5}^{(G_{u})}} - \braket{\mathbb{Z}_{9}^{(G_{u})}}/\sqrt{2}$.
  The total energy of the achiral D$_{\rm 3d}^{5}$ structure with $\delta u = 0$ is set as zero.
  }
  \end{center}
\end{figure*}

In the following, we explain each contribution of the SAMB to the Hamiltonian of the equilibrium Te and Se, which is summarized in Table~\ref{tab_SAMB_Te}.
The sign and magnitude of the coefficients of them in $H^{\rm SymCW}$ are given in Figs.~\ref{fig_samb_coeff}(a), (b) and (c), (d).

Let us first focus on the spinless terms in $H_{\rm CEF} + H_{\rm hop}^{\rm (b_{1})} + H_{\rm hop}^{\rm (b_{2})}$.
In $H_{\rm CEF}$, the E octupole $\mathbb{Z}_{1}^{(Q_{3\gamma})}$ represents the CEF in $xy$-plane, where the anisotropic potential at each atom is expressed by the E quadrupoles, $Q_{v}^{\rm (a)}$ and $Q_{xy}^{\rm (a)}$ as shown in Fig.~\ref{fig_samb_def}(c).
This term is active whenever the spatial inversion symmetry is lost.
On the other hand, $\mathbb{Z}_{2}^{(G_{u})}$ is the localized spinless ET quadrupole that is active only in chiral structure, and is described as the vortex-like alignment of the E quadrupoles at each atom as shown in Fig.~\ref{fig_samb_def}(e) (see Fig.~\ref{fig_ET}(d)).
The divergence of the vorticity corresponds to the handedness.
As summarized in Table~\ref{tab_SAMB_Te}, the weight of $\mathbb{Z}_{2}^{(G_{u})}$ in Te (Se), $\Delta_{2}^{(G_{u})} = 1.49\times 10^{-1}\, (2.04\times 10^{-1})$ [eV], 
is largest among the terms in $H_{\rm CEF}$ and is much larger than that of $\mathbb{Z}_{1}^{(Q_{3\gamma})}$, $\Delta_{1}^{(Q_{3\gamma})} = 5.71\times 10^{-2}\, (-6.36\times 10^{-2})$ [eV].

In $H_{\rm hop}^{\rm (b_{1})}$, the E monopole $\mathbb{Z}_{3}^{(Q_{0})}$ represents the NN spinless diagonal hopping between $p_{\mu}$ ($\mu = x,y,z$) orbitals at two neighboring atoms, as shown in Fig.~\ref{fig_samb_def}(b), while $\mathbb{Z}_{4}^{(Q_{3\gamma})}$ represents the NN off-diagonal hoppings between in-plane $p$ orbitals $p_{\mu}$ ($\mu = x,y$) at two neighboring atoms, as shown in Fig.~\ref{fig_samb_def}(d).
Moreover, $\mathbb{Z}_{5}^{(G_{u})}$ is the itinerant spinless ET quadrupole, i.e., the vortex-like alignment of E quadrupoles at the centers of the NN bonds, as shown in Fig.~\ref{fig_samb_def}(f), and the divergence of the vorticity corresponds to the handedness.
As summarized in Table~\ref{tab_SAMB_Te}, the weight of $\mathbb{Z}_{5}^{(G_{u})}$ in Te (Se), $t_{5}^{(G_{u})} = 6.76\, (9.35)$ [eV], is largest among the terms in $H_{\rm hop}^{\rm (b_{1})}$, and its value is much larger than that of $\mathbb{Z}_{3}^{(Q_{0})}$, $t_{3}^{(Q_{0})} = 1.93\, (2.13)$ [eV] and $\mathbb{Z}_{4}^{(Q_{3\gamma})}$, $t_{4}^{(Q_{3\gamma})} = 3.63\, (4.88)$ [eV].
Note that $\mathbb{Z}_{5}^{(G_{u})}$ represents the spin-independent off-diagonal real hopping between $p$ orbitals at NN atoms.
Specifically, as shown in Fig.~\ref{fig_samb_def}(g), for the hopping between B and C sites in the home unit-cell, $\mathbb{Z}_{5}^{(G_{u})}$ is given by
\begin{align}
t_{1} (c_{p_{z}, {\rm C}, \bm{0}}^{\dagger} c_{p_{y}, {\rm B}, \bm{0}}^{} + {\rm h.c.}),
\label{eq_Gu_b1a_2nd}
\end{align}
where $c_{p_{z}, {\rm C}, \bm{0}}^{\dagger}$ ($c_{p_{y}, {\rm B}, \bm{0}}^{}$) is the creation (annihilation) operator of the electron with $p_{z}$ ($p_{y}$) orbital at C (B) atom in the home unit-cell. 
Note that the NN off-diagonal hopping parameter $t_{1}$ of Te (Se) is related to $t_{5}^{(G_{u})}$ as $t_{1} = 1/(2\sqrt{3}) t_{5}^{(G_{u})} = 1.95\, (2.70)$ [eV], where the factor $1/\sqrt{3}$ appears by taking account of the number of the NN bonds.
Additionally, in the NNN hopping $H_{\rm hop}^{\rm (b_{2})}$, we can define the itinerant spinless ET quadrupole $\mathbb{Z}_{9}^{(G_{u})}$, as shown in Fig.~\ref{fig_samb_def}(h).
In Fig.~\ref{fig_samb_coeff}(a) and (c), the weight of $\mathbb{Z}_{9}^{(G_{u})}$ in Te (Se) is $t_{9}^{(G_{u})} = 4.10\, (3.11)$ [eV].
$\mathbb{Z}_{9}^{(G_{u})}$ represents the spin-independent off-diagonal real hopping between $p$ orbitals at NNN atoms.
For the hopping between A site in the home unit-cell and the C site in the NNN unit-cell with $\bm{R} = \bm{a}_{1} + \bm{a}_{2}$, $\mathbb{Z}_{9}^{(G_{u})}$ is given by
\begin{align}
t_{2} (c_{p_{z}, {\rm C}, \bm{R}}^{\dagger} c_{p_{y}, {\rm A}, \bm{0}}^{} + {\rm h.c.}).
\label{eq_Gu_b2a_2nd}
\end{align}
The NNN off-diagonal hopping parameter $t_{2}$ of Te (Se) is related to $t_{9}^{(G_{u})}$ as $t_{2} = 1/(2\sqrt{6}) t_{9}^{(G_{u})} = 0.837\, (0.634)$ [eV], where the factor $1/\sqrt{6}$ appears due to the number of the NNN bonds in the unit-cell.
Since the length of the NN bond is shorter than the NNN bond in chiral structures, the magnitude of $t_{1}$ is greater than $t_{2}$, while $t_{1} = t_{2}$ in the achiral D$_{\rm 3d}^{5}$ structure since the lengths of the NN and NNN bonds are the same.
In addition, since $\delta u$ of equilibrium Se is larger than that of Te, the ratio $t_{1}/t_{2}$ of Se is larger than that of Te, i.e., $t_{1}/t_{2} = 4.26\, ({\rm Se}) > 2.33\, ({\rm Te})$. 
It should be noted that we have confirmed that the magnitude of the coefficients decreases rapidly for further neighbor hoppings.

In $H_{\rm hop}^{\rm (b_{1})}$, there are another itinerant ET monopole and quadrupole, $\mathbb{Z}_{6}^{(G_{0})}$ and $\mathbb{Z}_{7}^{(G_{u})}$, which have the forms of $\bm{T}\cdot\bm{M}$ and $3T_{z}M_{z} - \bm{T}\cdot\bm{M}$. 
As shown in Fig.~\ref{fig_samb_def}(j), $\mathbb{Z}_{6}^{(G_{0})}$ represents the spin-independent off-diagonal real hopping between $p$ orbitals at NN atoms, while the anisotropic term $\mathbb{Z}_{7}^{(G_{u})}$ also appears in the trigonal helical structure.
Notably, the magnitude of the weight of $\mathbb{Z}_{7}^{(G_{u})}$ of Te and Se is greater than that of $\mathbb{Z}_{6}^{(G_{0})}$ owing to the anisotropy along the helical axis, while their values are much smaller than that of $\mathbb{Z}_{5}^{(G_{u})}$ and $\mathbb{Z}_{9}^{(G_{u})}$, as shown in Figs.~\ref{fig_samb_coeff}(a) and (c).

Next, we discuss the spinful terms in $H_{\rm SOC} + H_{\rm hop}^{\rm (b_{1})'}$.
In $H_{\rm SOC}$, there is the spinful E monopole $\mathbb{Z}_{10}^{(Q_{0})}\propto \bm{l}\cdot\bm{\sigma}$ representing the atomic SOC.
Moreover, there are the spinful ET monopole $\mathbb{Z}_{11}^{(G_{0})}\propto \bm{t}\cdot\bm{\sigma}$ and quadrupole $\mathbb{Z}_{12}^{(G_{u})}\propto 3t_{z} \sigma_{z} - \bm{t}\cdot\bm{\sigma}$ arising from the atomic SOC under chiral structure.
There is also the spinful ET quadrupole $\mathbb{Z}_{13}^{(G_{u})}$ having the same form as the spinless one $\mathbb{Z}_{2}^{(G_{u})}$, in which the spinless E quadrupoles are replaced with the spinful ones given by Eq.~(\ref{eq_Q2_sf_atomic}).
$\mathbb{Z}_{13}^{(G_{u})}$ is a kind of spinful CEF existing only in helical structure.
Additionally, $\mathbb{Z}_{14}^{(G_{0})}$ is the ET monopole having the flux structure of $\bm{G}^{\rm (a)'}$ at each atom, which corresponds to Fig.~\ref{fig_ET}(a) and schematically shown in Fig.~\ref{fig_samb_def}(k).
Note that however, the weights of these spinful multipoles except for $\mathbb{Z}_{10}^{(Q_{0})}$ in Te (Se) are much smaller than that of $\mathbb{Z}_{10}^{(Q_{0})}$, i.e., $\xi_{10}^{(Q_{0})} = 1.63\, (0.765)$ [eV].
Moreover, $\xi_{10}^{(Q_{0})} \, ({\rm Te})  > \xi_{10}^{(Q_{0})} \, ({\rm Se})$, since the SOC of the 5$p$ orbitals of Te is larger than that of 4$p$ orbitals of Se.

In $H_{\rm hop}^{\rm (b_{1})'}$, the spinful itinerant ET monopole $\mathbb{Z}_{15}^{(G_{0})}$ and quadrupole $\mathbb{Z}_{16}^{(G_{u})}$ are obtained by replacing the orbital angular momentum $\bm{l}$ with the spin angular momentum $\bm{\sigma}$ in $\mathbb{Z}_{6}^{(G_{0})}$ and $\mathbb{Z}_{7}^{(G_{u})}$.
As shown in Fig.~\ref{fig_samb_def}(j), $\mathbb{Z}_{15}^{(G_{0})}$ represents the spin-dependent imaginary hopping.
There also exists the anisotropic term $\mathbb{Z}_{16}^{(G_{u})}$ in the trigonal helical structure as mentioned before.
Eventually, the magnitude of these spinful multipoles in Te and Se is negligibly small as compared with the spinless itinerant ET monopole and quadrupole.

Here, it should be noted that the weight of $\mathbb{Z}_{15}^{(G_{0})}$ was evaluated as $\alpha_{15}^{(G_{0})} = 1.749$ eV in the previous study~\cite{RO_PRL_2022}, however, this was overestimated due to the lower precision of the previous method with the band dispersion fitting.

Additionally, the itinerant spinful ET quadrupole $\mathbb{Z}_{17}^{(G_{u})}$ in $H_{\rm hop}^{\rm (b_{1})'}$ has the same form as the spinless one $\mathbb{Z}_{5}^{(G_{u})}$, where the spinless E quadrupoles are replaced with the spinful ones given by Eq.~(\ref{eq_Q2_sf_atomic}).
$\mathbb{Z}_{17}^{(G_{u})}$ is the spin-dependent off-diagonal hopping between $p$ orbitals at NN sites.
Notably, as summarized in Table~\ref{tab_SAMB_Te}, the weight of $\mathbb{Z}_{17}^{(G_{u})}$ in Te and Se is largest among the terms in $H_{\rm hop}^{\rm (b_{1})'}$ and is much larger than that of $\mathbb{Z}_{15}^{(G_{0})}$ and $\mathbb{Z}_{16}^{(G_{u})}$.
$\mathbb{Z}_{18}^{(G_{0})}$ is the itinerant spinful ET monopole having the flux structure of $\bm{G}^{\rm (a)'}$ at each NN bond center as shown in Fig.~\ref{fig_samb_def}(l).
Its weight in Te is the second largest among the terms in $H_{\rm hop}^{\rm (b_{1})'}$, while the weight in Se is smaller than that of $\mathbb{Z}_{16}^{(G_{u})}$ because of the strong anisotropy along $z$ axis.
Anyway, the magnitude of them are negligibly small as compared with the itinerant spinless ET monopoles and quadrupoles.

From the above results, we conclude that the dominant contributions in $H^{\rm SymCW}$ to the equilibrium Te and Se arise from spinless ones; the localized ET quadrupole, $\mathbb{Z}_{2}^{(G_{u})}$, and the NN and NNN itinerant ET quadrupoles, $\mathbb{Z}_{5}^{(G_{u})}$ and $\mathbb{Z}_{9}^{(G_{u})}$, with the form of Eq.~(\ref{eq_Gu_classical}).

As discussed below, they also play a crucial role in stabilizing the helical structure.

\subsection{Electric toroidal quadrupole $G_{u}$ as a measure of chirality}
\label{sec:Gu}

In order to identify the predominant order parameter for structural chirality, let us discuss the evolution of each order parameter across the achiral-to-chiral structure change by changing $\tau\delta u$.
Since the ET monopole and quadrupole among SAMBs are the candidates of the predominant order parameters, we evaluate the expectation value of them in the achiral D$_{\rm 6h}^{1}$ ($\delta u = \pm 1/3$) and D$_{3d}^{5}$ ($\delta u = 0$) structures, and the right-handed ($-1/3 < \tau \delta u < 0$) and left-handed ($0 < \tau \delta u < 1/3$) chiral structures,
\begin{align}
\braket{\mathbb{Z}_{j}} = \frac{1}{N_{k}} \sum_{n\bm{k}} f(\epsilon_{n\bm{k}}^{\rm SymCW}) \braket{n\bm{k}|\mathbb{Z}_{j}| n\bm{k}},
\end{align}
where $f(\epsilon) = (e^{\beta \epsilon} + 1)^{-1}$ is the Fermi-Dirac distribution function ($\beta = 1/T$) and $\ket{n\bm{k}}$ is the $n$-th eigenvector of $H^{\rm SymCW}$ at $\bm{k}$ point.

Figures~\ref{fig_samb_coeff}(e) and (g) represent the $\tau \delta u$ dependence of the expectation value of the localized SAMBs in Te and Se, respectively, where the temperature $T$ is fixed at $0.02$ eV.
Figures~\ref{fig_samb_coeff}(e) and (g) indicate that the spinless localized ET quadrupole $\braket{\mathbb{Z}_{2}^{(G_{u})}}$ is zero in the achiral structures, while $\braket{\mathbb{Z}_{2}^{(G_{u})}} \neq 0$ in the chiral structures and its value increases linearly with increasing $\tau\delta u$ from the achiral D$_{\rm 3d}^{5}$ structure.
Additionally, its sign is reversed depending on the chirality $\tau = \pm 1$.
Thus, $\braket{\mathbb{Z}_{2}^{(G_{u})}}$ plays a proper indicator of chirality, and its sign corresponds to the handedness of chiral structure.
On the other hand, the spinless localized E octupole $\braket{\mathbb{Z}_{1}^{(Q_{3\gamma})}}$ and the spinful localized E monopole $\braket{\mathbb{Z}_{10}^{(Q_{0})}}$ are even function of $\tau\delta u$.
Since the SOC of the 5$p$ orbitals of Te is larger than that of 4$p$ orbitals of Se, $\braket{\mathbb{Z}_{10}^{(Q_{0})}}$ of Te is larger than that of Se.

Figures~\ref{fig_samb_coeff}(f) and (h) represent the $\tau \delta u$ dependence of the expectation value of the itinerant SAMBs in Te and Se, respectively.
As shown in Figs.~\ref{fig_samb_coeff}(f) and (h), the NN and NNN itinerant spinless ET quadrupoles, $\braket{\mathbb{Z}_{5}^{(G_{u})}}$ and $\braket{\mathbb{Z}_{9}^{(G_{u})}}/\sqrt{2}$, become identical when $\delta u = 0$, since the length of the NN and NNN bonds are the same.
On the other hand, the difference of them,
\begin{align}
\phi = \braket{\mathbb{Z}_{5}^{(G_{u})}} - \frac{1}{\sqrt{2}}\braket{\mathbb{Z}_{9}^{(G_{u})}},
\label{eq_net_Gu}
\end{align}
becomes nonzero when the system is chiral, and its sign corresponds to the handedness.
Here, the coefficient $1/\sqrt{2}$ of $\braket{\mathbb{Z}_{9}^{(G_{u})}}$ is taken into account, since the number of the NNN bonds in the unit-cell is twice the NN bonds.
Figures~\ref{fig_samb_coeff}(i) and (k) represent the change of the net itinerant spinless ET quadrupole $\phi$ as a function of $\tau \delta u$.
As shown in Figs.~\ref{fig_samb_coeff}(i) and (k), $\phi$ increases linearly with increase of $\tau\delta u$ from the achiral D$_{\rm 3d}^{5}$ structure, and its value is largest among the ET multipoles. 
On the other hand, as similar to the localized E multipoles, the spinless E monopole $\braket{\mathbb{Z}_{3}^{(Q_{0})}}$ and octupole $\braket{\mathbb{Z}_{4}^{(Q_{3\gamma})}}$ are even function of $\tau\delta u$.
Therefore, we conclude that $\phi$ plays the predominant order parameter for structural chirality of Te and Se.

Moreover, the NN itinerant spinless ET quadrupole $\mathbb{Z}_{5}^{(G_{u})}$ in Eq.~(\ref{eq_net_Gu}) plays a crucial role to form the strong covalent bond among NN atoms in the unit-cell, which contributes to the energy gain through $E_{\rm one-e}$ in Eq.~(\ref{eq_E_total}) to stabilize the helical structure.
In other words, $\phi$ becomes finite for $0 < \delta u < 0.06533$ $(0.1013)$ in Te (Se), and therefore, the chiral crystal structure becomes energetically stable, because the energy gain from the kinetic energy exceeds the energy loss from Coulomb interactions between ions and between electrons.
On the other hand, as $\delta u$ approaches $1/3$, the system becomes unstable because the energy loss due to the Coulomb interactions between ions and electrons exceeds the energy gain from the kinetic energy. 
It should be noted that in realizing the helical crystal structure of Te and Se, the presence of orbital degrees of freedom is significant, otherwise the orbital-exchange hoppings corresponding to the itinerant spinless ET quadrupoles $\mathbb{Z}_{5}^{(G_{u})}$ and $\mathbb{Z}_{9}^{(G_{u})}$ do not exist, and no kinetic energy gain is obtained.

In order to further confirm that $\phi$ is the predominant order parameter for structural chirality of Te and Se, we present the change of the total energy as a function of $\phi$, as shown in Figs.~\ref{fig_samb_coeff}(j) and (l). 
The total energy as a function of $\phi$ reaches an extremum when $\phi$ coincides with its equilibrium value $\phi^{*}$, indicating clearly that $\phi$ plays a role of the order parameter for structural chirality, and stabilizing the helical structures of Te and Se.

\subsection{$G_{u}$ as an origin of monopole-like OAM texture in the momentum space}
\label{sec:OAM}

\begin{figure*}[t]
  \begin{center}
  \includegraphics[width=0.9\hsize]{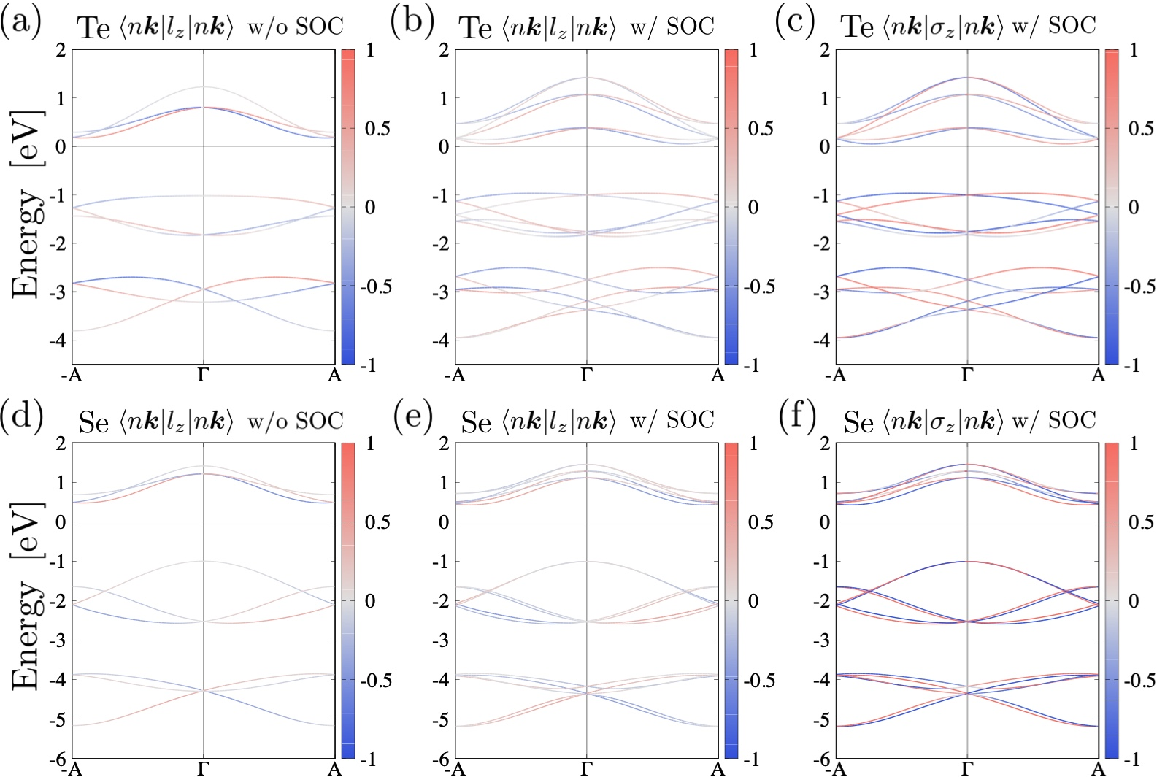}\vspace{8mm}
  \caption{
  \label{fig_orbital}
  Orbital and spin angular momentum textures along -A$-\Gamma-$A line of (a)-(c) Te and (d)-(f) Se obtained from the SymCW model (a), (d) without SOC and (b), (c), (e), (f) with SOC.
  The colormap represents the $z$ component of (a),(b),(d),(e) the orbital polarization $\braket{n\bm{k}|l_{z}|n\bm{k}}$ and (c),(f) the spin polarization $\braket{n\bm{k}|\sigma_{z}|n\bm{k}}$.
  The Fermi energy is set as the origin.
  }
  \end{center}
\end{figure*}

In this section, we show that the itinerant spinless ET quadrupole $\mathbb{Z}_{5}^{(G_{u})}$ as discussed in the previous subsection is also the leading origin of the monopole-like OAM texture in the momentum space.

In the chiral structure of Te and Se, the monopole-like antisymmetric spin and orbital splittings, $\braket{n\bm{k}|M_{z}|n\bm{k}} = -\braket{n -\bm{k}|M_{z}|n -\bm{k}}$ ($M_{z} = \sigma_{z}, l_{z}$), appear as shown in Figs.\ref{fig_orbital}(a)-(c) and (d)-(f), respectively.
It should be emphasized that the large orbital splitting with $\sim 1.0$ [eV] appears irrespective of SOC as shown in Figs.~\ref{fig_orbital}(a) and (d).
Since the SOC of Se is much smaller than that of Te [See Figs.~\ref{fig_samb_coeff}(b) and (d)], the spin splitting of Se with $\sim 0.1$ [eV] is much smaller than that of Te, as shown in Figs.~\ref{fig_orbital}(c) and (f).
Therefore, the orbital splitting of Se is more prominent than that of Te.

Next, in the framework of the systematic analysis method proposed in Refs.~\cite{SH_PRB_2020, RO_JPSJ_2022}, we extract the important parameters in the SymCW model leading to the OAM texture by evaluating the following quatity,
\begin{align}
&\Gamma^{i}(\bm{k}) = \mathrm{Tr}\left[l_{z} (H^{\rm SymCW}(\bm{k}))^{i}\right],
\label{eq_Gamma_k}
\end{align}
where $(H^{\rm SymCW}(\bm{k}))^{i}$ is the $i$-th power of the Hamiltonian matrices of the SymCW model.
For simplicity, we consider only the dominant SAMBs up to the NN bonds and ignore $\mathbb{Z}_{6}^{(G_{0})}$, $\mathbb{Z}_{7}^{(G_{u})}$, and $H_{\rm SOC}$ in Eq.~(\ref{eq_SymCW_2}).
We obtain the lowest-order nontrivial contribution as
\begin{align}
&\Gamma^{3}(\bm{k}) \propto \tau t_{3}^{(Q_{0})} \left[ (t_{5}^{(G_{u})})^{2} - 2 (t_{4}^{(Q_{3\gamma})})^{2} \right] \sin \left(k_{z}\right).
\end{align}
This result indicates that the dominant contribution to the monopole-like OAM texture arises from the itinerant spinless E ocrupole $\mathbb{Z}_{4}^{(Q_{3\gamma})}$ and ET quadrupole $\mathbb{Z}_{5}^{(G_{u})}$.
In particular, in the equilibrium structure of Te and Se, $t_{5}^{(G_{u})}$ is about twice as large as $t_{4}^{(Q_{3\gamma})}$, as shown in Figs.~\ref{fig_samb_coeff}(a) and (c), suggesting that the former is the leading origin of such OAM texture.

It is emphasized that, as similar to the observations in the cubic B20 family~\cite{Yang_OAM_2023, Brinkman_OAM_2024, Yen_OAM_2024}, the monopole-like OAM texture in the momentum space is expected to be observed in trigonal chiral materials such as Te and Se via CD signals in soft x-ray photoemission spectroscopy measurements.
As discussed above, the monopole-like OAM texture in Te and Se is induced by the itinerant spinless ET quadrupole, whereas in the cubic B20 family, it must be related to the ET monopoles or the fourth-rank of ET multipole as there is no monoaxial principal axis.
Although the CD signals in photoemission arise from both orbital and spin splitting, the orbital contribution would dominate over spin one especially in Se because of its relatively week SOC.

\section{Conclusion}
\label{sec:conclusion}

In summary, we have demonstrated that for the typical chiral crystals Te and Se, the itinerant spinless electric toroidal quadrupole $G_{u}$, representing the spin-independent real orbital-exchange hopping between $(p_{x},p_{y})$ and $p_{z}$, plays a significant role both in stabilizing the helical structure and providing monopole-like orbital angular momentum texture in momentum space.

First, in the framework of the symmetry-adapted multipole basis (SAMB), we have clarified that the order parameter of chirality in non-cubic systems is described by not only the ET monopole $G_{0}$ but also the ET quadrupole of ($3z^{2}-r^{2}$)-type $G_{u}$, which belong to the totally symmetric irreducible representation in monoaxial chiral point groups. 
Next, using the symmetry-adapted closest Wannier model constructed from the DFT calculations for the trigonal Te and Se, we have shown that the essential part of the Hamiltonian is indeed expressed by the spinless and spinful $G_{0}$ and $G_{u}$ SAMBs. 

Then, we have evaluated the expectation values of these spinless and spinful $G_{0}$ and $G_{u}$ as well as the other irrelevant achiral components, across the achiral-to-chiral structure change.
Our results have clearly demonstrated that the expectation value of the itinerant spinless $G_{u}$ is the predominant contribution to the essential aspects of chirality, and its sign is reversed depending on the crystal chirality.
The itinerant spinless $G_{u}$ leads to the strong covalent bond between nearest-neighbor atoms in the unit-cell and stabilizes the helical structure.
Moreover, the difference between the itinerant spinless $G_{u}$ for the nearest-neighbor and next-nearest-neighbor bonds, as called $\phi$ in the main text, plays a role of the predominant order parameter for structural chirality of Te and Se. 
The same $G_{u}$ gives rise to the monopole-like orbital angular momentum texture in the momentum space as well, which can be observed by the circular-dichroism in soft x-ray photoemission spectroscopy measurement.

The present analysis procedure can be applied to other chiral crystals, such as trigonal $\alpha$-HgS~\cite{ishitoTrulyChiralPhonons2023}, $M$Si$_{2}$ ($M$ = Nb, Ta)~\cite{Shiota_CISS_PRL_2021, Shishido_CISS_ARL_2021}.
The predominant order parameter in the cubic B20 family~\cite{Yang_OAM_2023, Brinkman_OAM_2024, Yen_OAM_2024}, such as PdGa, PtGa, and CoSi, is expected to be the ET monopole or ET hexadecapole, whose investigation is left for future study.
Our findings highlight the importance of orbital degrees of freedom in chiral materials both in stabilizing chiral crystal structures and resulting band structures accompanied with considerably large orbital angular momentum spliting.
It sheds light on a deeper understanding of chirality from the viewpoint of the interplay between electronic and structural aspects.

\section{Acknowledgement}
\label{sec:acknowledgement}
The authors thank Akane Inda, Shingo Kuniyoshi, Satoru Hayami, Yuki Yanagi, Takayuki Ishitobi, Kazumasa Hattori, Ming-Chun Jiang, Hikaru Watanabe, Takuya Nomoto, Ryotaro Arita, Yoshihiko Togawa, Yusuke Kato, and Junichiro Kishine, for fruitful discussions.
This work was done under the Special Postdoctoral Researcher Program at RIKEN and supported by JSPS KAKENHI Grants Numbers JP23K03288, JP23H00091, and the grants of Special Project (IMS program 23IMS1101), and OML Project (NINS program No, OML012301) by the National Institutes of Natural Sciences.

\bibliography{ref}

\end{document}